\documentclass[lettersize,journal,final]{IEEEtran}
\usepackage{amsmath,amsfonts}
\usepackage{algorithm}
\usepackage{array}
\usepackage[caption=false,font=normalsize,labelfont=sf,textfont=sf]{subfig}
\usepackage{textcomp}
\usepackage{stfloats}
\usepackage{url}
\usepackage{verbatim}
\usepackage{graphicx}
\usepackage{cite}
\usepackage{booktabs,bm,color,multirow}
\usepackage[export]{adjustbox}
\usepackage{algpseudocode,cases,enumitem}
\usepackage{amssymb}

\algrenewcommand\algorithmicindent{1.0em}

\algrenewcommand\algorithmicindent{1.0em}

\newcommand{\mat}[1] {\mathbf{#1}}
\newcommand{\vect}[1] {\mathbf{#1}}

\newcommand{\tabcell}[1]{\begin{tabular}{@{}c@{}}#1\end{tabular}}
\newcommand*\rot{\rotatebox[origin=c]{90}}

\newcommand{\R}[1] {\mathbb{R}^{#1}}

\def\numero{%
  {\rm N}%
  \raise1.05ex\hbox{\fiverm \b{o}}%
}

\newtheorem{thm}{Theorem}

\DeclareMathOperator*{\argminA}{argmin}
\DeclareMathOperator*{\argmax}{argmax}

\begin{document}

\title{FROST-BRDF: A Fast and Robust Optimal Sampling Technique for BRDF Acquisition}

\author{Ehsan Miandji, Tanaboon Tongbuasirilai, Saghi Hajisharif, Behnaz Kavoosighafi and Jonas Unger        
\IEEEcompsocitemizethanks{ 
\IEEEcompsocthanksitem The authors are with Link\"{o}ping University, Sweden, except Tanaboon Tongbuasirilai who is with Kasetsart University, Thailand. 
\IEEEcompsocthanksitem Email:\{ehsan.miandji\}@liu.se.
\IEEEcompsocthanksitem Email:\{tanaboon.to\}@ku.th
\IEEEcompsocthanksitem Email:\{saghi.hajisharif,behnaz.kavoosighafi,jonas.unger\}@liu.se}
\thanks{Manuscript received Month XX, 20XX; revised Month XX 20XX.}}

\markboth{Submitted to IEEE Transactions on Visualization and Computer Graphics}
{Shell \MakeLowercase{\textit{Miandji et al.}}: FROST-BRDF: A Fast and Robust Optimal Sampling Technique for BRDF Acquisition}


\maketitle

\begin{abstract}
 Efficient and accurate BRDF acquisition of real world materials is a challenging research problem that requires sampling millions of incident light and viewing directions. To accelerate the acquisition process, one needs to find a minimal set of sampling directions such that the recovery of the full BRDF is accurate and robust given such samples. In this paper, we formulate BRDF acquisition as a compressed sensing problem, where the sensing operator is one that performs sub-sampling of the BRDF signal according to a set of optimal sample directions. To solve this problem, we propose the Fast and Robust Optimal Sampling Technique (FROST) for designing a provably optimal sub-sampling operator that places light-view samples such that the recovery error is minimized. FROST casts the problem of designing an optimal sub-sampling operator for compressed sensing into a sparse representation formulation under the Multiple Measurement Vector (MMV) signal model. The proposed reformulation is exact, i.e. without any approximations, hence it converts an intractable combinatorial problem into one that can be solved with standard optimization techniques. As a result, FROST is accompanied by strong theoretical guarantees from the field of compressed sensing. We perform a thorough analysis of FROST-BRDF using a $10$-fold cross-validation with publicly available BRDF datasets and show significant advantages compared to the state-of-the-art with respect to reconstruction quality. Finally, FROST is simple, both conceptually and in terms of implementation, it produces consistent results at each run, and it is at least two orders of magnitude faster than the prior art. 
\end{abstract}

\section{Introduction}  \label{sec:introduction}

Current applications in computer graphics and sensor simulation, e.g., predictive rendering and generation of synthetic data for machine learning, put ever increasing demands on physical accuracy and realism. A fundamental enabler for accurate rendering and sensor simulation is accurate modeling of material scattering behaviours, generally described by the Bidirectional Reflectance Distribution Function (BRDF), \cite{Nicodemus:1992}. The need for physical accuracy has put appearance capture and measurements of real world BRDFs at the heart of many production pipelines. Efficient and accurate acquisition of real world materials is a challenging, and yet to be solved research problem,~\cite{ward92, Dana99, Matusik2003:Model, Lensch:2005, Ghosh10}. The key challenges are the sheer number of samples that need to be measured and the complexity of the measurement setups required to record the data to model a BRDF. 


The space of existing methods for BRDF acquisition is extensive, ranging from approximate methods using deep learning with the goal of instant acquisition \cite{kalantari-svbrdf,shi-svbrdf,deschaintre2018single}, to dense and accurate measurements using gonio-reflectometers that may require several hours or even days for capturing a single material. The goal in this paper is to minimize the acquisition time without sacrificing the quality, i.e., to utilize only a small number of measurements, e.g. 10 to 40, to accurately reconstruct a densely sampled BRDF. Given the number of samples, the main challenge here is to find the optimal sample locations, i.e., placements of the light and the sensor, such that the BRDF recovery error is kept minimal. This problem is inherently combinatorial and often intractable. For instance, a BRDF in Rusinkiewicz's parameterization \cite{Rusinkiewicz98} has $90\times90\times180=1,458,000$ elements, hence taking $20$ samples without replacement produces a set of $\big(\begin{smallmatrix}  1,458,000 \\  20\end{smallmatrix}\big)\approx 10^{104}$ possibilities for sample placement. Our proposed method solves such an intractable problem by reformulating it into a form that can be efficiently solved using standard optimization techniques while providing theoretical guarantees.  

In this paper, we present a novel compressed sensing formulation of the BRDF acquisition problem. Particularly, we show that the problem of finding the optimal light and camera sample directions can be seen as designing a sub-sampling operator. We then propose the Fast and Robust Optimal Sampling Technique (FROST) for designing a provably optimal BRDF sub-sampling operator. FROST is a non-parametric learning algorithm that takes as input the sparse representation of a set of BRDFs, i.e. the training set, and computes a small set of optimal sample directions to be used by, e.g., gonioreflectometers. Once the optimal samples are acquired, the reconstruction of the full BRDF is carried out by solving a least squares problem. The key observation in designing FROST is that by reformulating the intractable combinatorial problem of designing a sub-sampling operator into a sparse representation problem under the Multiple Measurement Vector (MMV) signal model, standard and well-established sparse recovery algorithm can be used to solve the problem. Since the reformulation is exact, i.e. no approximation is made, the solution of the MMV sparse recovery algorithm is the optimal sub-sampling operator. Moreover, existing theoretical guarantees for MMV signal recovery can be used to derive theoretical guarantees for finding optimal BRDF sampling directions. 

FROST is deterministic with theoretical guarantees, and produces optimal sampling directions at each run. It is important to emphasize that FROST is general in that the basis/dictionary and reconstruction algorithm can be chosen freely. Although there are many different options, we use a PCA-basis and an $\ell_2$ reconstruction algorithm in our experiments and evaluations. The reason for this is that it conforms to previous methods that we compare to and enables us to clearly demonstrate the strengths and improved quality introduced by FROST itself.

\noindent The main contributions of this paper can be summarized as:

\begin{itemize}
    \item{A novel formulation of the BRDF acquisition problem using compressed sensing.}
    \item{A novel sensing matrix design algorithm, abbreviated FROST, for designing a provably optimal sub-sampling operator of arbitrary signals.}
    \item{A novel data-driven application of FROST for computing the optimal set of directions given a training set for BRDF acquisition.}
    \item{Comparisons to the state-of-the-art, showing significant improvements in terms of reconstruction error with equal number of samples, as well as computational complexity.}
\end{itemize}

We evaluate our method using the measured BRDFs from MERL \cite{Matusik2003:MERL} and RGL-EPFL \cite{Dupuy:EPFL} BRDF databases. We use a $10$-fold cross-validation to thoroughly analyze the performance of FROST-BRDF by randomly dividing the data sets into multiple sets of training and testing sets. Our experiments demonstrate significant advantages in terms of reconstruction error and image quality when compared to the method of Nielsen et al.\cite{Nielsen2015}. Moreover, FROST-BRDF produces consistent results at each run due to deterministic optimal sample placements. To further demonstrate the robustness of FROST-BRDF, we report results for the DTU dataset \cite{Nielsen2015}, without including any BRDF from this dataset in the training set.
 
The paper is organized as follows. We present the related works in Section \ref{sec:background}. Readers who are not familiar with the fields of sparse representation and compressed sensing can refer to sections 1 and 2 in the supplementary material, where we briefly discuss relevant concepts used in the derivation of FROST-BRDF. In Section \ref{sec:method}, we present our approach, starting by the formulation of the problem using compressed sensing, and its solution via our proposed method, FROST. Implementation details, i.e. the choice of the dictionary and the reconstruction algorithm are presented in Section \ref{sec:impl}. The results are summarized in Section \ref{sec:result}, followed by a discussion around limitations and future work directions in Section \ref{sec:limitation}. Finally, we conclude this paper in Section \ref{sec:conclusion}. Code and data for this paper are available at \url{www.dummy.url}.


\section{Related work}   \label{sec:background}

\noindent\textbf{BRDF measurement.} \quad \label{sec:background:brdf_measurement}
A commonly used technique for acquiring BRDFs is gonioreflectometry, that measures reflectance on a flat sample \cite{Nicodemus:1992, Guarnera:2016}. However, this method is labor-intensive and requires a large number of images for a high-resolution BRDF acquisition. Image-based measurement techniques, such as Marschner et al. \cite{Marschner00}, have been proposed to alleviate such limitations. This method, later adopted by Matusik et al. \cite{Matusik2003:MERL} for isotropic BRDF measurement, resulted in the publicly available MERL database for the research community. More recently, Dupuy and Jakob \cite{Dupuy:EPFL} proposed an adaptive parameterization for adaptive BRDF measurements. These advances have led to the development of various complex acquisition setups for a wide range of surfaces, e.g. \cite{ward92, Dana99, Ghosh10, Gardner03, Tunwattanapong13}.


\noindent\textbf{Sparse BRDF sampling.} \quad
Matusik et al. \cite{Matusik2003:Model} proposed a sparse sampling method based on a linear combination of BRDFs with 45 principal components using PCA. However, this method requires up to 800 samples to achieve accurate reconstructed BRDFs. On the other hand, the behavior of the bivariate representations found in BRDF parameterizations suggests that fewer measurements may be sufficient. Tongbuasirilai et al. \cite{Tongbuasirilai.2019.CompactAI} proposed a sparse measurement method based on two factors, requiring only 180 samples, that was primarily designed for glossy materials. 
Romeiro et al. \cite{Romeiro08, Romeiro10} also presented BRDF estimation techniques using bivariate parameterization from a single image. Aittala et al. \cite{aittala2015two} restricted the surface appearance to a class of texture-like materials and enabled two-shot capturing and recovering of SVBRDFs. The advancements in the field of deep learning have led to the development of multiple methods for estimating BRDFs from images \cite{Chen20, boss2021nerd, boss19, kalantari-svbrdf, GAO19}. Researchers have explored techniques for instant capturing and estimation of SVBRDFs using a single image by identifying an appropriate procedural model from a library \cite{shi-svbrdf}, and a small set of images captured with a flashlight via generative adversarial networks \cite{guo2020materialgan}, or convolutional neural networks \cite{li2017modeling,li2018materials,deschaintre2018single,xu2018deep}, and autoencoders \cite{kang2018efficient,kang2019learning}. While these methods have shown visually plausible results, physical accuracy of the recovered materials may not be guaranteed. 

Nielsen et al. \cite{Nielsen2015} proposed a heuristic discrete gradient descent algorithm to minimize the condition number of a dictionary computed on a training set. This technique, however, has major shortcomings. First, the iterative optimization technique utilizes random initializations at each iteration, hence converging to a different local minima at each run. Second, we observe that minimizing the condition number does not necessarily correlate with minimizing the recovery error. Finally, due to the heuristic approach used, the obtained sampling directions are not necessarily optimal. Consequently, deriving theoretical guarantees becomes challenging, if not impossible. The method proposed by Xu et al. \cite{Xu2016:sampling} improves upon Nielsen et al.'s approach by changing the error metric from the condition number to the reconstruction error and noise. However, this method still suffers from random initialization and inconsistent results for each run due to the heuristic optimization technique utilized. Their proposed method is designed for two-shot near-field acquisition, hence it cannot be directly compared with FROST-BRDF. Yu et al. \cite{svbrdf-ravi} extended the method of Nielsen et al. to find optimal sampling directions for spatially-varying BRDF acquisition. This was achieved by removing grazing angles from the optimization, aligning multiple viewpoints, and correcting for near-field effects.

\noindent\textbf{Data-driven BRDF model.} \quad
Sparse BRDF acquisition is closely related to the data-driven BRDF models, as existing methods decompose a BRDF into a dictionary and a set of coefficients, which are the inputs to our proposed FROST-BRDF technique. 
Factorization-based models assume that the measured BRDFs can be factorized in various ways to enable efficient representations \cite{Lawrence:2004, soler18:CGF, Bilgili11, Tongbuasirilai.2019.CompactAI}, that can be used for e.g. interactive editing \cite{Kautz99} and real-time rendering \cite{Ben-Artzi06}. Efficient BRDF representations have been explored previously, e.g. for embedding of measured BRDFs with editing capabilities on a BRDF manifold \cite{soler18:CGF}, using diffuse-specular separation to reduce the coefficients for encoding measured BRDFs \cite{sun2018connect}, and using special weighting functions for non-parametric models \cite{Bagher16}. Sztrajman et al. \cite{sztrajman2021neural} proposed a neural network for the compact representation of BRDFs. Later, Tongbuasirilai et al. \cite{Tongbuasirilai22} proposed a dictionary learning technique to represent measured BRDFs on a manifold constructed from an ensemble of multidimensional dictionaries. Exploring the utilization of data-driven BRDF models introduced above in FROST-BRDF is an interesting research direction that we would like to pursue in the future. 

\noindent\textbf{Compressive sensing.} \quad 
Over the past decade, compressive sensing has been adopted by the Graphics community for various applications, such as rapidly capturing light transport data \cite{peers2009compressive,sen2009compressive}, compressive rendering \cite{ehsan-eg2015,comp-rendering}, light field photography \cite{marwah2013compressive,saghi-lfv}, geometry processing \cite{cs-geometry}, and hyperspectral imaging \cite{cs-hyperspectral}. Moreover, several studies have explored the reconstruction of BRDFs from a set of random sparse measurements using compressive sensing techniques, including Fourier domain measurements \cite{Seylan2013brdf,Zupancic2013sparse}, learned multidimensional dictionaries \cite{Miandji:Thesis,ehsan-tog2019}, and block Discrete Cosine Transform (DCT) \cite{Otani2019}. However, to the best of our knowledge, sensing operator design for optimal and deterministic compressive acquisition of BRDF data has not been explored.

\begin{figure*}
\centering
\includegraphics[width=0.9\linewidth]{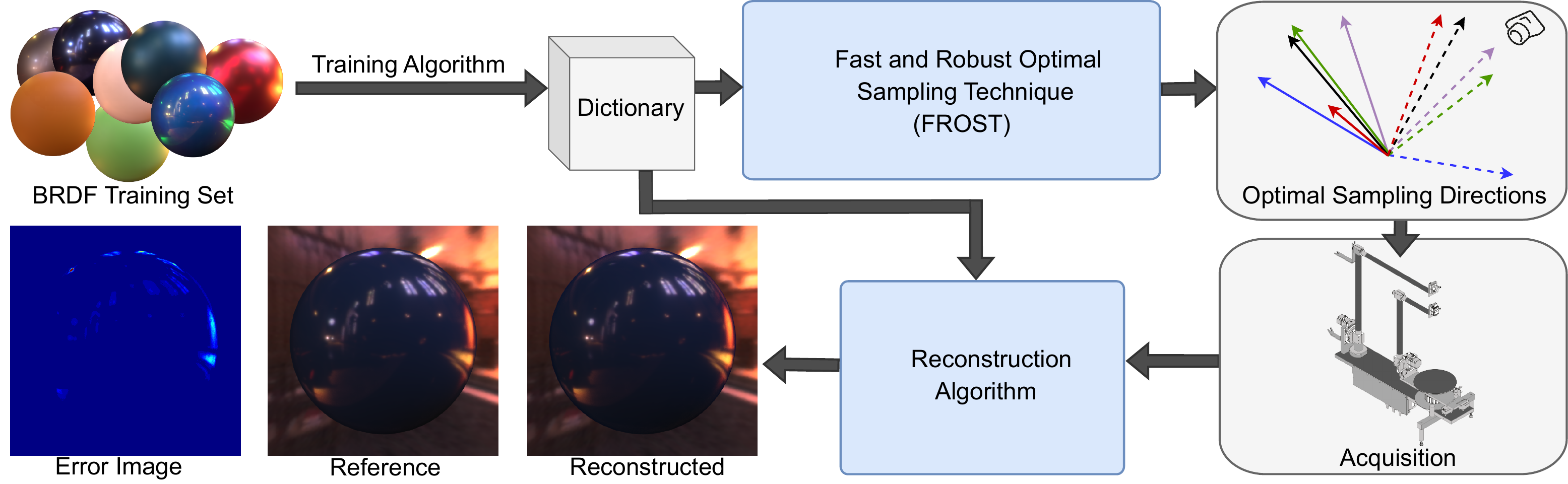}
\caption{shows an overview of our proposed technique, FROST-BRDF, for sparse BRDF acquisition. A dictionary is trained given a BRDF training set. We then utilize FROST, a novel algorithm for computing an optimal sub-sampling operator corresponding to a small number of light-camera sampling directions. After obtaining the samples we reconstruct the full BRDF using the dictionary. The example BRDF shown here is \textit{specular-blue-phenolic}, which was not included in the training set, and was reconstructed using only $10$ optimally placed samples out of $90\times90\times180$ elements, i.e. the resolution of the full BRDF.}
\label{fig:Pipeline}
\end{figure*}

\section{Sparse BRDF acquisition}   \label{sec:method}

In this section, we present our method for BRDF acquisition using a small set of optimally-placed samples, see Figure \ref{fig:Pipeline}. 
We assume that the reader is familiar with fundamentals of sparse representation and compressed sensing. In particular, FROST relies on the concepts of Single Measurement Vector (SMV) and MMV signal models, as well as sensing matrix design and dictionary learning. For convenience, we present these topics in the supplementary document accompanying this manuscript. To evaluate our proposed method, we focus on homogeneous surfaces and isotropic materials. 

{\noindent\textbf{Notations- }}Throughout the paper, we adopt the following notational convention. Vectors and matrices are represented by boldface lower-case ($\vect{a}$) and bold-face upper-case letters ($\mat{A}$), respectively. A finite set of objects is indexed by superscripts, e.g. $\left\{\mat{A}^{(i)}\right\}_{i=1}^{N}$, whereas individual elements of $\vect{a}$ and $\mat{A}$ are denoted $\vect{a}_i$ and $\mat{A}_{i_1,i_2}$, respectively. The $i$th column in $\mat{A}$ is denoted $\mat{A}_i$, while the $i$th row is denoted $\mat{A}_{i,:}$. We use the notation $\mat{A}_J$ to represent the columns of $\mat{A}$ that are in the index set $J$. The $\ell_p$ norm of a vector $\vect{s}$, for $1\le p\le \infty$, is denoted by $\|\vect{s}\|_p$. Frobenius norm is denoted $\|\vect{s}\|_F$. The $\ell_0$ pseudo-norm of a vector, denoted $\|\vect{s}\|_0$, defines the number of non-zero elements in the vector. The $\ell_0$ pseudo-norm defined over the rows of a matrix $\mat{S}$ is denoted $\|\mat{S}\|_{r0}$, which counts the number of nonzero rows of $\mat{S}$.


%
%


\subsection{Problem formulation} \label{sec:method:formulation}

Our goal in this paper is to find a small set of optimally-placed measurements, where each measurement corresponds to a pair of light and camera directions, and the optimality condition is defined as the BRDF reconstruction error. As shown in Figure \ref{fig:Pipeline}, FROST-BRDF contains three stages: 1. a learning-based dictionary that admits a sparse representation of a set of BRDFs, 2. FROST, an algorithm for computing a minimal set of optimal sampling directions given the learned dictionary, and 3. a reconstruction algorithm that recovers a full BRDF from its measurements. The contributions of this paper lie at the second stage, i.e. the FROST algorithm for finding the optimal sample directions. However, for clarity, we start by formulating the problem of recovering a full BRDF from its measurements using compressed sensing. 

We assume that the BRDF, $\vect{b}\in\R{n}$, we would like to estimate is sparse under a dictionary $\mat{D}\in\R{n\times k}$. This means that $\vect{b}=\mat{D}\vect{s}$, where $\vect{s}\in\R{k}$ is sparse, satisfying $\|\vect{s}\|_0<\tau$ for a user-defined sparsity parameter $\tau$. Let $\mat{\Phi}\in\R{m\times n}$, $m<n$, be a sub-sampling operator that constructs a linear measurement model as $\vect{y} = \mat{\Phi}\vect{b}$. Note that, a sub-sampling operator is a special case of a sensing operator where the operator selects predefined elements from the input signal $\vect{b}$. This operator can be realized by constructing an $n\times n$ identity matrix, keeping $m$ rows corresponding to the sample locations. The column indices of nonzero values in $\mat{\Phi}$ define the sample locations, i.e. the elements in $\vect{b}$ to be sampled. 

With this formulation and given the sub-sampling operator $\mat{\Phi}$, the dictionary $\mat{D}$, and the measurements $\vect{y}$, we solve the following optimization problem to recover the full BRDF from its measurements
\begin{equation} \label{eq:cs-recon}
    \hat{\vect{s}} = \argminA_{\vect{s}} \|{\vect{s}}\| \;\;\mathrm{s.t.}\;\; \|{\vect{y} - \mat{\Phi} \mat{D} \vect{s}}\|^2_2 \le \epsilon,
\end{equation}
where $\epsilon$ is a user-defined threshold for reconstruction error. Once the sparse coefficients, $\hat{\vect{s}}$, are obtained, the full BRDF is obtained by evaluating $\hat{\vect{b}}=\mat{D}\hat{\vect{s}}$. It can be noted from \eqref{eq:cs-recon} that we have not specified a specific type of norm for the term $\|{\vect{s}}\|$. This is because the type of norm depends on the choice of the dictionary, e.g. whether the dictionary is orthonormal or overcomplete. More details on our choice of the norm will be presented in Section \ref{sec:impl}. 



Given \eqref{eq:cs-recon}, the next step is to design an optimal sub-sampling operator, $\mat{\Phi}$. The optimality condition can be defined in two ways: (a) given the number of samples the optimal sub-sampling operator minimizes the measurement error, or (b) given a reconstruction error threshold the optimal sub-sampling operator minimizes the number of samples. The problem of computing an optimal measurement matrix given a dictionary, or a set of training signals, is known as sensing operator design \cite{sensing-optim}. Depending on the application, one typically wants to enforce a certain structure on the sensing matrix, e.g., optimal circulant matrices as proposed in \cite{circulant-sensing} for applications where the measurement model involves a convolution. For BRDF acquisition, however, the structure of the sensing matrix is simple, as described above, since it corresponds to the sub-sampling of a discrete signal. We can formulate the problem of sensing matrix design for sub-sampling as follows
\begin{align} 
    \argminA_{\mat{\Phi} \in \Psi} g(\mat{\Phi}) \;\;\mathrm{s.t.}\;\; \|{\vect{y} - \mat{\Phi} \mat{D} \vect{s}}\|^2_2 \le \epsilon, \label{eq:sensing-design-point-g}  \\
    \argminA_{\mat{\Phi} \in \Psi} \|{\vect{y} - \mat{\Phi} \mat{D} \vect{s}}\|^2_2 \;\;\mathrm{s.t.}\;\;  g(\mat{\Phi}) = m,   \label{eq:sensing-design-point-l2}
\end{align}
where $\Psi$ is the set of all sub-sampling operators, $\mat{\Phi}\in\mathbb{R}^{m\times n}$, and the function $g(.)$ defines the number of measurements, i.e. the number of rows in $\mat{\Phi}$. 

Using \eqref{eq:sensing-design-point-g} and \eqref{eq:sensing-design-point-l2} for the design of a sub-sampling operator, $\mat{\Phi}$, assumes that $\mat{D}$ and $\vect{s}$ are known. However, in practice we do not have access to $\mat{D}$ and $\vect{s}$, since they have to be computed from the BRDF we would like to acquire. To tackle this problem, we pre-compute $\mat{D}$ and $\vect{s}$ given a training set of BRDFs. Define $\mat{T}\in\R{n\times t}$ as a training set of $t$ BRDFs arranged as columns. Assuming that $\mat{T}$ has a sparse representation in a dictionary $\mat{D}$, we have 
\begin{equation}    \label{eq:sparse-rep-train}
    \mat{T}=\mat{D}\mat{S}+\mat{\delta}, \; \mathrm{where} \; \|\mat{S}_i\|_0 \le\tau, \; \forall i \in \{1,\dots,t\}, 
\end{equation}
where $\mat{\delta}$ is the error introduced by the sparse representation; i.e. when the signal is compressible rather than having an exact sparsity. Each column of $\mat{S}$ in \eqref{eq:sparse-rep-train} contains the coefficients of the corresponding BRDF in $\mat{T}$. Note that it is not required that all $\mat{S}_i$ have the same sparsity. However, without loss of generality, we impose this condition to simplify the exposition. Then, finding the optimal sub-sampling operator for the training set $\mat{T}$ is formulated as
\begin{align} 
    \argminA_{\mat{\Phi} \in \Psi} g(\mat{\Phi}) \;\;\mathrm{s.t.}\;\; \|{\vect{Y} - \mat{\Phi} \mat{D} \vect{S}}\|^2_2 \le \epsilon, \label{eq:sensing-design-point-T-g}  \\
    \argminA_{\mat{\Phi} \in \Psi} \|{\vect{Y} - \mat{\Phi} \mat{D} \vect{S}}\|^2_2 \;\;\mathrm{s.t.}\;\;  g(\mat{\Phi}) = m,   \label{eq:sensing-design-point-T-l2}
\end{align}
where, with a slight abuse of notation, $\mat{Y}=\mat{\Phi}\mat{T}$. The error threshold, $\epsilon$, is user-defined, but can also be computed from $\mat{\delta}$ and the measurement noise. To facilitate comparisons with previous work, we mainly focus on \eqref{eq:sensing-design-point-T-l2}, where the number of samples is user-defined. However, as described in Section \ref{sec:method:frost}, both equations can be solved using our proposed method with a minor modification. 
 
Due to the high-dimensional nature of measured BRDFs, an exhaustive search through all the sensing matrices within $\Psi$ is not possible for solving \eqref{eq:sensing-design-point-T-l2}. For instance, the set of sub-sampling operators, $\Psi$, considering $20$ samples from a BRDF with a resolution of $90\times90\times180$, contains ${1458000\choose 20} \approx 10^{104}$ possible candidates for the optimal sub-sampling operator.

\subsection{Solution via FROST}  \label{sec:method:frost}


Given the combinatorial nature of the BRDF sub-sampling problem, defined in \eqref{eq:sensing-design-point-T-l2}, our main goal in designing FROST is to cast the problem into a form that can be solved using standard optimization techniques.
Let us start by considering the term $\mat{\Phi}\mat{D}\vect{s}$ in \eqref{eq:cs-recon}. The main objective in the sparse representation of a signal is to find the least number of atoms in a dictionary, i.e. the least number of columns in $\mat{D}$, that best approximate the given signal. The key observation is that the problem of sparse representation can, on one hand, be seen as a sub-sampling problem defined over the atoms of $\mat{D}$. On the other hand, the sub-sampling operator $\mat{\Phi}$ samples the rows of $\mat{D}$. Therefore, we can intuitively say that the operator $\mat{\Phi}$ can be viewed as a column sampling operator for the inverse of the dictionary, $\mat{D}^{-1}$. If we multiply both sides of \eqref{eq:sparse-rep-train} by $\mat{D}^{-1}$, we obtain
\begin{equation} \label{eq:dict-coeff-inv}
\mat{S} = \mat{D}^{-1}\mat{T} - \mat{D}^{-1}\delta.
\end{equation}
To this end, we ignore the term containing $\delta$ since we assume that the representation error is negligible. Seen as a least squares problem, Equation \eqref{eq:dict-coeff-inv} implies that $\mat{D}^{-1}$ and $\mat{S}$ are the known variables, while $\mat{T}$ is the unknown. Such formulation is often referred to as the \emph{analysis} model \cite{rubinstein-analysis}, while the formulation in \eqref{eq:sparse-rep-train} is referred to as the \emph{synthesis} model. We can now formulate the analysis variants of \eqref{eq:sensing-design-point-T-g} and \eqref{eq:sensing-design-point-T-l2} as 
\begin{align} 
    \vect{\lambda} &= \argminA_{\mat{Z}} \;\; \|\mat{Z}\|_{r0} \;\;\mathrm{s.t.}\;\; \left\|\mat{S} - \mat{D}^{-1}\mat{Z}\right\|^2_2 \le \epsilon, \;\text{and} \label{eq:frost-r0} \\
    \vect{\lambda} &= \argminA_{\mat{Z}} \;\; \left\|\mat{S} - \mat{D}^{-1}\mat{Z}\right\|^2_2 \;\;\mathrm{s.t.}\;\;  \|\mat{Z}\|_{r0} = m, \label{eq:frost-l2}
\end{align}
respectively, where $\|.\|_{r0}$ is the $\ell_0$ norm defined over rows, i.e. the number of nonzero rows. Equations \eqref{eq:frost-r0} and \eqref{eq:frost-l2} are classical examples of sparse representation under the MMV model, where we assume that the same support is shared among all signals. The variable $\mat{Z}$ is a replacement for the training set $\mat{T}$, which is the unknown according to the analysis formulation in \eqref{eq:dict-coeff-inv}. However, since we are only interested in optimal sample locations for the training set, i.e. the rows of $\mat{T}$, we do not need to recover the matrix $\mat{Z}$, but rather the row support of $\mat{Z}$. We denote the row support of $\mat{Z}$ by $\lambda$. 
The set $\lambda$ defines the optimal sample locations for the training set $\mat{T}$. Note that while the sub-sampling operator $\mat{\Phi}$ does not explicitly appear in \eqref{eq:frost-r0} and \eqref{eq:frost-l2}, it is implicitly defined using the support set $\lambda$. In other words, one can construct a sub-sampling operator, from an identity matrix as described above, using the support set $\lambda$. This demonstrates the equivalence of \eqref{eq:sensing-design-point-T-g} and \eqref{eq:sensing-design-point-T-l2} with \eqref{eq:frost-r0} and \eqref{eq:frost-l2}, respectively. The index set $\lambda$ can be used in, e.g., a gonioreflectometer to obtain a set of BRDF samples for recovering the full BRDF by solving \eqref{eq:cs-recon}. More details on the reconstruction algorithm and the choice of the dictionary $\mat{D}$ will be presented in Section \ref{sec:impl}.

FROST reformulates the problem of designing an optimal sub-sampling operator under the synthesis model as a support recovery problem for a set of signals using the MMV model in the analysis domain. Therefore, in essence, FROST determines which rows in the training set $\mat{T}$ carry the highest amount of information such that by only having access to these sample locations, we can recover any other unseen BRDF. From a different perspective, since FROST is defined over the analysis domain, we can say that FROST finds the most important columns of $\mat{D}^{-1}$ such that the representation error $\|\mat{S}-\left(\mat{D}^{-1}\right)_{\lambda}\mat{T}_{\lambda,.}\|_2^2$ is minimized. Hence, we are seeking a row sampling of $\mat{T}$, or equivalently a column sampling of $\mat{D}^{-1}$, such that we can closely match the sparse coefficient matrix $\mat{S}$. A key aspect of FROST is the equivalence of \eqref{eq:frost-r0} and \eqref{eq:frost-l2} with \eqref{eq:sensing-design-point-T-g} and \eqref{eq:sensing-design-point-T-l2}, respectively. In other words, the global minimizers of \eqref{eq:frost-r0} and \eqref{eq:frost-l2}, are also the global minimizers of \eqref{eq:sensing-design-point-T-g} and \eqref{eq:sensing-design-point-T-l2}, respectively.

\begin{algorithm}[t]   
	\caption{Simultaneous Orthogonal Matching Pursuit (SOMP) for solving Equation \eqref{eq:frost-r0} or \eqref{eq:frost-l2}} \label{alg:somp}
	\begin{algorithmic}[1]
        \Require{BRDF coefficients $\mat{S} \in\mathbb{R}^{p\times t}$, BRDF dictionary inverse $\mat{D}^{-1} \in\mathbb{R}^{p\times n}$, and user-defined number of samples $m \ge 1$ or an error threshold $\epsilon$} 
        \Ensure{Support set $\lambda$ (i.e. sample locations)}
        \State $l \gets 1$
        \State $\lambda^{(0)} \gets \varnothing$ \Comment{The support set at iteration 0}
        \State $\mat{R}^{(0)} \gets \mat{S}$    \Comment{The residual at iteration 0}
        \While{$\|\mat{R}\|_2\ge\epsilon$ or $l < m$} 
             
             \Comment{Eq. \eqref{eq:frost-r0} or \eqref{eq:frost-l2}, respectively}
              
              \State $j \gets \underset{i}{\argmax}  \left\| (\mat{D}^{-1})^{T}_{i} \mat{R}^{(l-1)} \right\|_{1}$    \Comment{Atom selection}
		     \State $\lambda^{(l)} \gets \lambda^{(l-1)} \cup \{j\}$  \Comment{Support set update}
		     \State $\mat{R}^{(l)} \gets \mat{S}-(\mat{D}^{-1}_{\lambda^{(l)}}) (\mat{D}^{-1}_{\lambda^{(l)}})^{\dagger} \mat{S}$     \Comment{Residual update}
		     \State $l \gets l + 1$
		\EndWhile
	\end{algorithmic}
\end{algorithm}

Since FROST solves an MMV support recovery problem, a large number of algorithms can be used to solve \eqref{eq:frost-r0} or \eqref{eq:frost-l2} \cite{Tropp:SOMP,Kim:SSMP,Han:Orthogonal_Subspace}. Such algorithms are often accompanied by theoretical performance guarantees; i.e. conditions under which the solution obtained by the algorithm corresponds to the global minimizer of \eqref{eq:frost-r0} or \eqref{eq:frost-l2}. While the accuracy and performance of each algorithm varies depending on signal and dictionary properties, in this paper we utilize the most simple algorithm, in terms of implementation, namely Simultaneous Orthogonal Matching Pursuit (SOMP) introduced by Tropp et al. \cite{Tropp:SOMP}. SOMP is an adaptation of Orthogonal Matching Pursuit (OMP), which is a greedy algorithm for the SMV model with theoretical guarantees \cite{omp-spl,tropp-greed}, for solving the MMV model. Algorithm \ref{alg:somp} presents an implementation of SOMP for solving \eqref{eq:frost-r0} or \eqref{eq:frost-l2}. 

A thorough theoretical analysis of FROST is outside the scope of this paper, and it is left for future work. However, we provide an example of such performance guarantees by rewriting the main result reported by Tropp et al. \cite{Tropp:SOMP} according to our formulation in \eqref{eq:frost-l2}. The theorem utilizes the cumulative mutual coherence for a dictionary, denoted $\mu_{1}(m)$, which measures the maximum total correlation between a fixed atom and $m$ distinct atoms, see \cite{Tropp:SOMP} for more details.
\begin{thm}
    Let $\mat{T}_{\alpha,0}$ be the optimal solution of \eqref{eq:frost-l2} according to the true support set $\alpha$. Moreover, let $\mat{T}_{\lambda,0}$ be the solution of SOMP after $m$ iterations, i.e. $|\alpha|=|\lambda|=m$. Assume that $\mu_{1}(m)<1/2$, then
    \begin{multline}
        \left\|\mat{S}-\left(\mat{D}^{-1}\right)_{\lambda}\mat{T}_{\lambda,0}\right\|_F \le \\ 
        \sqrt{1+mt\frac{1-\mu_{1}(m)}{[1-2\mu_{1}(m)]^2}} \left\|\mat{S}-\left(\mat{D}^{-1}\right)_{\alpha}\mat{T}_{\alpha,0}\right\|_F.
    \end{multline}
    \label{thm:relation}
\end{thm}
Theorem \ref{thm:relation} above states the relation between the number of samples, $m$, and the number of training BRDFs, $t$, to the approximation error.

According to the discussion above, it is evident that FROST does not rely on the choice of dictionary, and that both learning based and analytical dictionaries may be used for $\mat{D}$ to solve \eqref{eq:frost-r0} or \eqref{eq:frost-l2}. It can be argued that FROST relies on the inverse of the dictionary, i.e. $\mat{D}^{-1}$, which means that the dictionary has to be orthogonal. However, even if we are given an overcomplete dictionary, $\mat{D}\in\R{n\times k}$, where $k>n$, we can compute the Moore–Penrose inverse of $\mat{D}$ as $\mat{D}^{\dagger}=\mat{D}^T(\mat{D}\mat{D}^T)^{-1}$. Since overcomplete dictionary learning algorithms produce dictionaries with linearly independent rows, the pseudo-inverse exists. Accordingly, we solve the following optimization problem instead of \eqref{eq:frost-l2}:
\begin{equation} \label{eq:frost-l2-pseudo}
    \vect{\lambda} = \argminA_{\mat{Z}} \;\; \left\|\mat{S} - \mat{D}^{\dagger}\mat{Z}\right\|^2_2 \;\;\mathrm{s.t.}\;\;  \|\mat{Z}\|_{r0} = m. 
\end{equation}
A thorough analysis of FROST-BRDF using overcomplete \cite{ksvd} or multidimensional dictionary ensembles \cite{ehsan-tog2019} is left for future work.

\section{Implementation} \label{sec:impl}
FROST can be implemented in a number of ways, depending on the choice of dictionary and reconstruction algorithm. In choosing suitable algorithms for these two components, our main criterion was a fair comparison with the state-of-the-art. We compare our method with that of Nielsen et al. \cite{Nielsen2015} as the state-of-the-art algorithm for recovering BRDFs from a small set of optimally placed measurements. Therefore, we utilize the same dictionary and reconstruction algorithm as in \cite{Nielsen2015}. 
Note that the only requirement in choosing a dictionary for our method is that the dictionary used for FROST and the reconstruction algorithm are the same. The dictionary used in our implementation is described in Section \ref{sec:impl:dict}, while the reconstruction algorithm is detailed in Section \ref{sec:impl:recon}.



The formulation of FROST-BRDF uses a vectorized form of a measured BRDF. For instance, considering a MERL BRDF \cite{Matusik2003:MERL} with Rusinkiewicz's parameterization \cite{Rusinkiewicz98}, the length of $\vect{b}$ is $n=90\times90\times180=1458000$. Accordingly, we have $\mat{D}\in\mathbb{R}^{1458000\times k}$, where $k$ is the number of atoms. Note that we remove the invalid angles for each BRDF to avoid placing samples on them. Due to the high dynamic range presented in measured BRDFs, linear BRDFs have been shown to perform poorly for fitting to parametric and nonparametric models \cite{Matusik2003:Model, Low2012, Nielsen2015}. To tackle this issue, the set of training BRDFs are transformed using the \textit{log-relative mapping} introduced in \cite{Nielsen2015}.  

\subsection{Dictionary} \label{sec:impl:dict}

We use Principal Component Analysis (PCA) to construct an orthogonal dictionary with $k$ atoms, i.e. the principal components. 

\begin{figure}[ht]
    \begin{center}
    \includegraphics[width=1.0\linewidth]{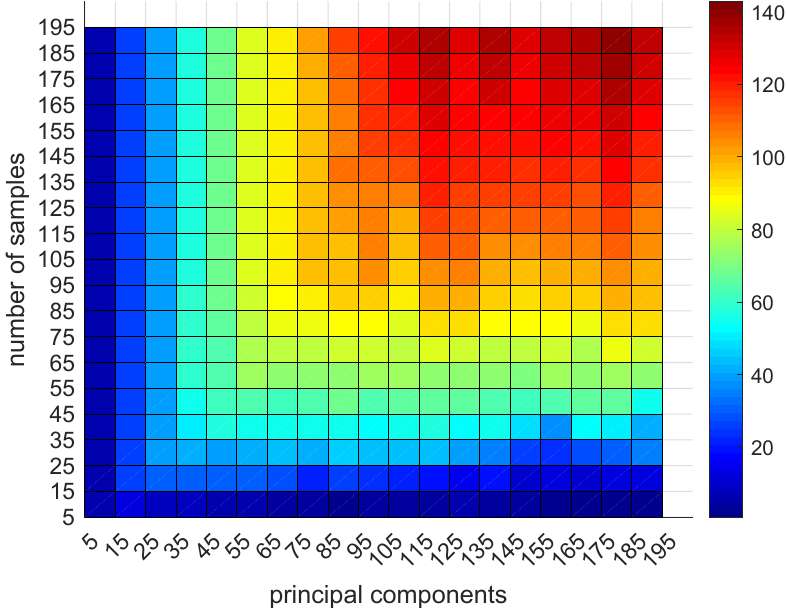}
    \end{center}
    \caption{Inverse MSE with respect to the number of samples, $m$, and the number of principal components, $p$.}
    \label{fig:p-m}
\end{figure}

Let $\mat{T} \in \mathbb{R}^{n \times t}$ be a matrix containing the set of $t$ vectorized training BRDFs as columns. After computing the mean of all BRDFs in $\mat{T}$ and subtracting the result from $\mat{T}$, we perform a truncated Singular Value Decomposition (SVD) as $\mat{T} = \mat{U}\mat{\Sigma}\mat{V}^{T}$, with $k<t$ singular values, where $\mat{U}\in\mathbb{R}^{n \times k}$ and $\mat{V}\in\mathbb{R}^{t \times k}$ contain the left and right singular vectors, respectively. The diagonal matrix $\mat{\Sigma} \in \mathbb{R}^{k \times k}$ holds the singular values. Define $\mat{D}=\mat{U}\mat{\Sigma}$ as the dictionary and $\mat{S}=\mat{V}^{T}$ as the set of coefficients. The number of principal components and its effect on BRDF model accuracy has been extensively studied \cite{sun2018connect}. Since our method is not sensitive to this parameter, see Figure \ref{fig:p-m}, we fixed this parameter to the number of samples, i.e. we set $m=k$. The same approach is used by the state-of-the-art method we compare to \cite{Nielsen2015}. Note that changing $k$ does not require us to recompute the SVD, but rather we can choose the first $k$ columns from $\mat{D}$ and $\mat{V}$. 
\subsection{Reconstruction algorithm}    \label{sec:impl:recon}

The reconstruction algorithm is independent of the algorithm used for obtaining the optimal sample directions. Once the sample directions are obtained, recovering the full BRDF is a least squares problem, which can be regularized using a suitable norm, see Eq. \eqref{eq:cs-recon}. The choice of the norm depends on the dictionary used for the representation. For instance, using a PCA dictionary, an $\ell_2$ regularizer is adequate, as noted by Nielsen et al. \cite{Nielsen2015}, which enables a closed-form solution with fast reconstruction when compared to an $\ell_0$ or $\ell_1$ solution. As a result, we utilize an $\ell_2$ regularizer for the reconstruction to facilitate comparisons with \cite{Nielsen2015}. Even though we utilize the same dictionary and reconstruction algorithm, our results in Section \ref{sec:result} show significant advantages for FROST in terms of quality of reconstruction and speed. This is due to the optimal sample placements by FROST.

Let $\vect{b}_{\lambda}$ be the measurements according to the FROST optimal sample locations obtained from an unseen BRDF, i.e. a BRDF in the testing set. Due to the orthogonality of the PCA dictionary, the reconstruction is simply a least squares problem, which we solve using ridge regression \cite{blanz-ridge}
\begin{equation}    \label{eq:recon} 
    \argminA_{\vect{s}} \; \|\vect{b}_{\lambda} - \mat{D}_{\lambda,:}\vect{s}\|_2^2 + \eta\|\vect{s}\|_2,
\end{equation}
where $\eta$ is a user-defined parameter. Note that since $\mat{D}$ is a PCA dictionary, which has sorted atoms based on importance, it is adequate to utilize the first $m$ columns of $\mat{D}_{\lambda,:}$ when solving \eqref{eq:recon}. This is an observation made in \cite{Nielsen2015}. Therefore, for a fair comparison with the method of Nielsen et al. \cite{Nielsen2015}, we set $m=k$, i.e. we replace $\mat{D}_{\lambda,:}$ with $\mat{D}_{\lambda,1:m}$ in \eqref{eq:recon}. We also set the same value for the parameter $\eta$, i.e. $\eta=40$. The coefficients obtained by solving \eqref{eq:recon} are then used to compute the recovered full BRDF as $\hat{\vect{b}}=\mat{D}\vect{s}$. When $m<k$, the matrix $\mat{D}_{\lambda,:}$ becomes overcomplete. In this case, we need to replace the $\ell_2$ regularizer in \eqref{eq:recon} with an $\ell_0$ or $\ell_1$ norm.





\begin{table*}[ht]
    \centering
    \caption{Optimal sampling locations from our approach, FROST, and Nielsen et al. \cite{Nielsen2015}. We performed 5 trials for Nielsen et al. \cite{Nielsen2015}, due to the inconsistency of the outcomes between each run. We use the MERL BRDF coordinate system.}
    \small
    \begin{tabular}{|l|ccc|ccc|ccc|ccc|ccc|ccc|} 
        \cline{2-19}
        \multicolumn{1}{l|}{\multirow{3}{*}{}} & \multicolumn{3}{c|}{\multirow{2}{*}{FROST-BRDF}} & \multicolumn{15}{c|}{Nielsen et al. \cite{Nielsen2015}} \\ 
        \cline{5-19}
        \multicolumn{1}{l|}{} & \multicolumn{3}{c|}{} & \multicolumn{3}{c|}{Trial 1}  & \multicolumn{3}{c|}{Trial 2} & \multicolumn{3}{c|}{Trial 3} & \multicolumn{3}{c|}{Trial 4} & \multicolumn{3}{c|}{Trial 5} \\ 
        \cline{2-19}
        \multicolumn{1}{l|}{} & \multicolumn{1}{l}{$\theta_h$} & \multicolumn{1}{l}{$\theta_d$} & \multicolumn{1}{l|}{$\phi_d$} & \multicolumn{1}{l}{$\theta_h$} & \multicolumn{1}{l}{$\theta_d$} & \multicolumn{1}{l|}{$\phi_d$} & \multicolumn{1}{l}{$\theta_h$} & \multicolumn{1}{l}{$\theta_d$} & \multicolumn{1}{l|}{$\phi_d$} & \multicolumn{1}{l}{$\theta_h$} & \multicolumn{1}{l}{$\theta_d$} & \multicolumn{1}{l|}{$\phi_d$} & \multicolumn{1}{l}{$\theta_h$} & \multicolumn{1}{l}{$\theta_d$} & \multicolumn{1}{l|}{$\phi_d$} & \multicolumn{1}{l}{$\theta_h$} & \multicolumn{1}{l}{$\theta_d$} & \multicolumn{1}{l|}{$\phi_d$}  \\ 
        \hline
        \multirow{5}{*}{\rotatebox{270}{$m=5$}} & 8 & 55 & 111 & 2 & 88 & 53 & 5 & 81 & 22 & 2 & 88 & 21 & 5 & 87 & 161 & 0 & 82 & 145 \\
         & 50 & 29 & 145 & 6 & 0 & 11 & 7 & 11 & 86 & 6 & 2 & 86 & 7 & 0 & 4 & 6 & 7 & 85 \\
         & 57 & 78 & 79 & 14 & 71 & 75 & 21 & 72 & 85 & 14 & 68 & 105 & 14 & 72 & 86 & 21 & 72 & 85 \\
         & 61 & 86 & 95 & 43 & 81 & 114 & 43 & 81 & 115 & 43 & 81 & 114 & 43 & 81 & 114 & 43 & 81 & 115 \\
         & 61 & 88 & 88 & 68 & 0 & 166 & 63 & 18 & 154 & 66 & 1 & 179 & 72 & 7 & 178 & 72 & 4 & 147 \\
        \hline
    \end{tabular}    
    \label{tab:our_dtu_5samples}
\end{table*}

\begin{figure}[ht]
    \begin{center}
    \includegraphics[width=0.95\linewidth]{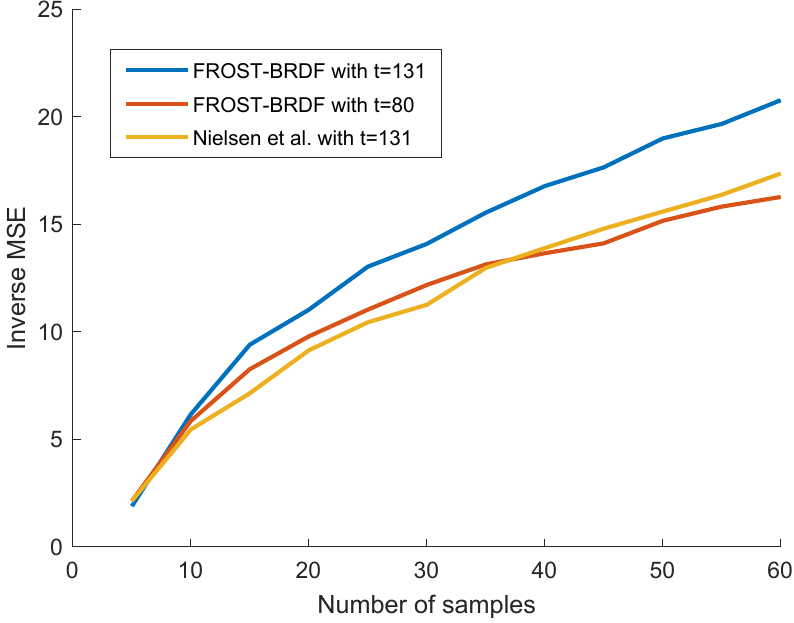}
    \end{center}
    \caption{\label{fig:dtu_vs_our} Inverse MSE of BRDF reconstruction with respect to the number of samples obtained from FROST-BRDF (our method) with 131 materials (blue), and 80 materials (red) in the training set, compared with Nielsen et al. \cite{Nielsen2015} with 131 materials in the training set (yellow).}
\end{figure}

\begin{figure*}[ht]
    \centering
    \setlength{\tabcolsep}{0.001cm}
    \begin{tabular}{ccccccc}
        \includegraphics[width=0.135\linewidth]{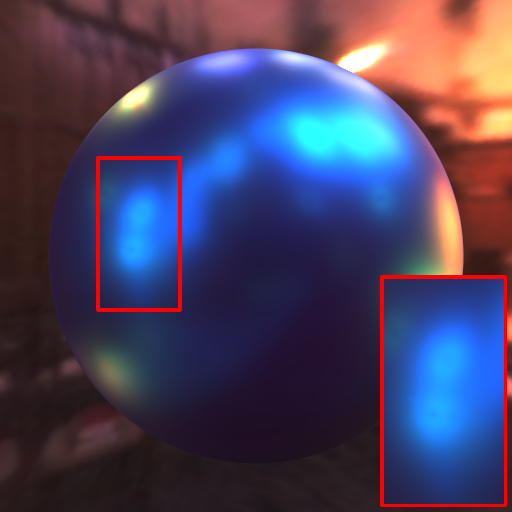} &
        \includegraphics[width=0.135\linewidth]{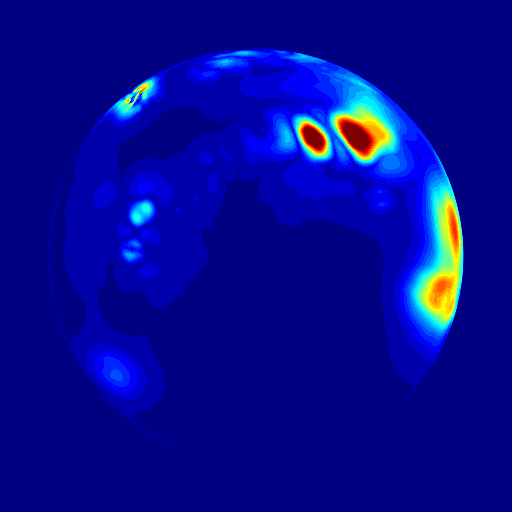} &
        \includegraphics[width=0.135\linewidth]{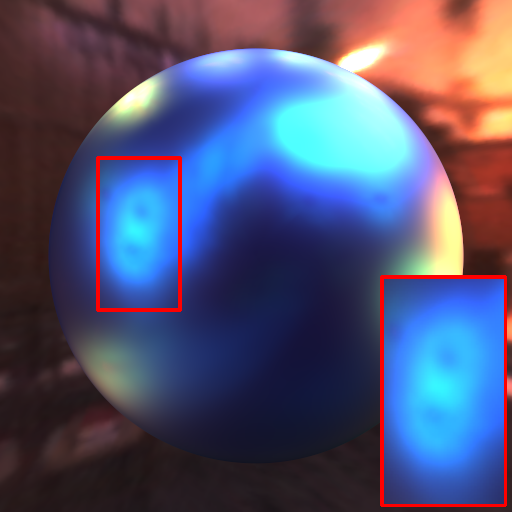} &
        \includegraphics[width=0.135\linewidth]{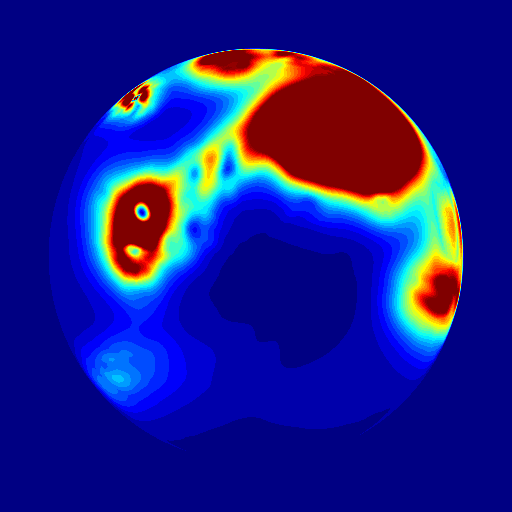} &
        \includegraphics[width=0.135\linewidth]{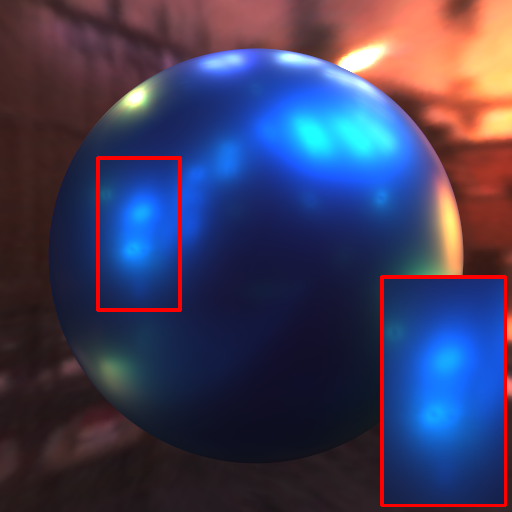} &
        \includegraphics[width=0.135\linewidth]{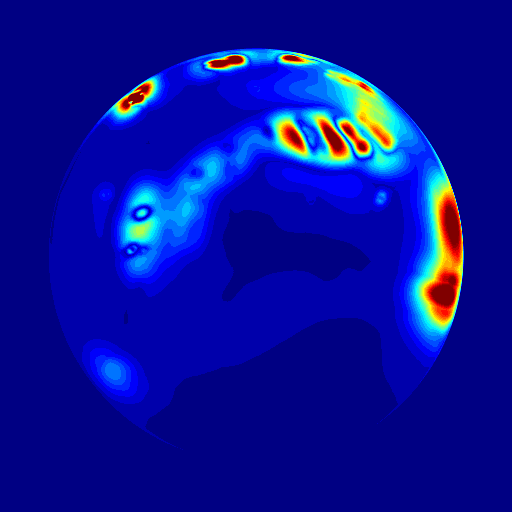} &
        \includegraphics[width=0.135\linewidth]{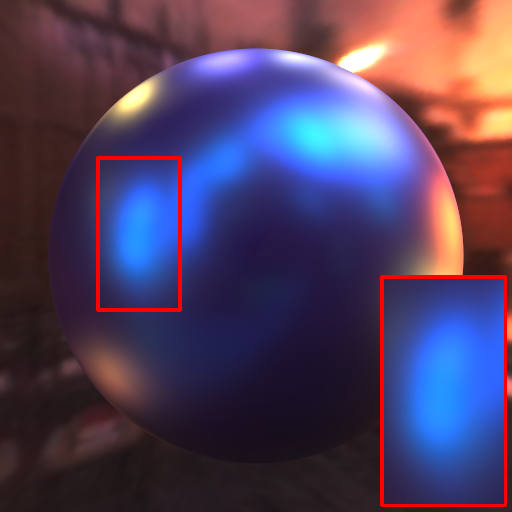}
        \\
        \multicolumn{2}{c}{Nielsen-Trial 1} & \multicolumn{2}{c}{Nielsen-Trial 2} & \multicolumn{2}{c}{Nielsen-Trial 3} & Reference 
        \\ 
        \includegraphics[width=0.135\linewidth]{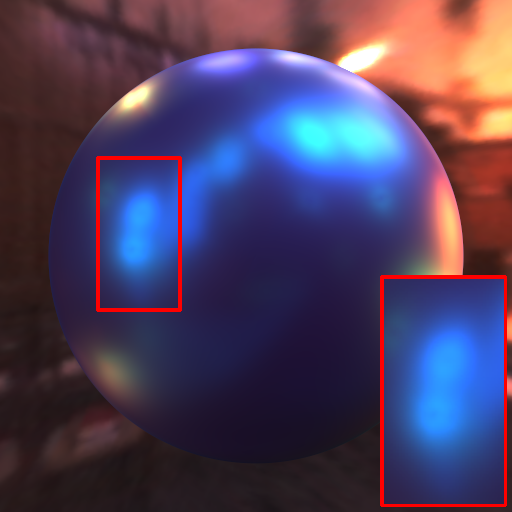} &
        \includegraphics[width=0.135\linewidth]{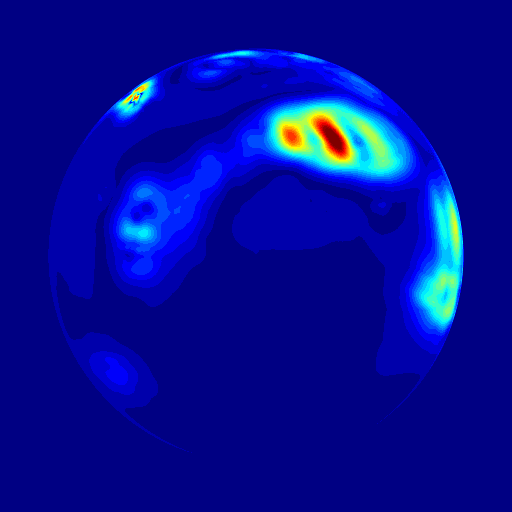} &
        \includegraphics[width=0.135\linewidth]{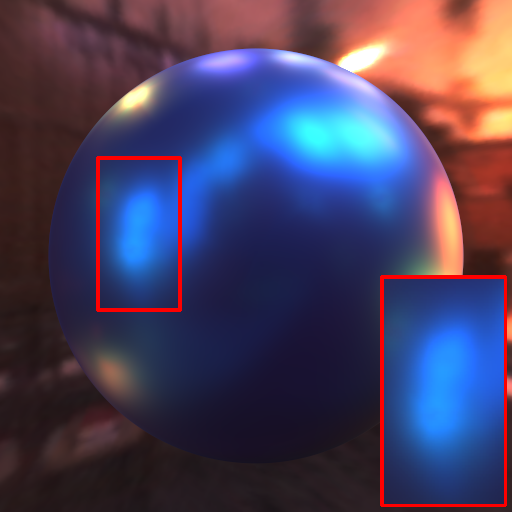} &
        \includegraphics[width=0.135\linewidth]{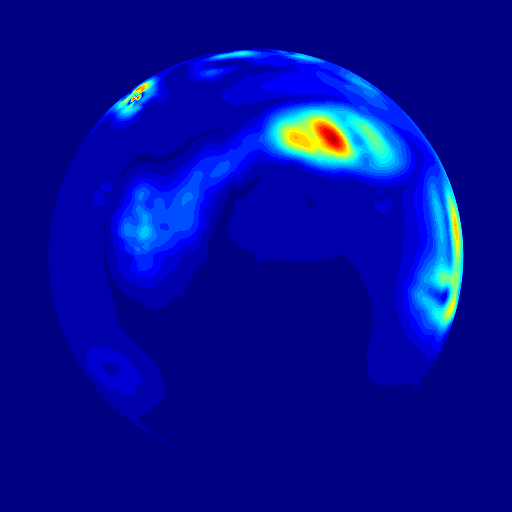} &
        \includegraphics[width=0.135\linewidth]{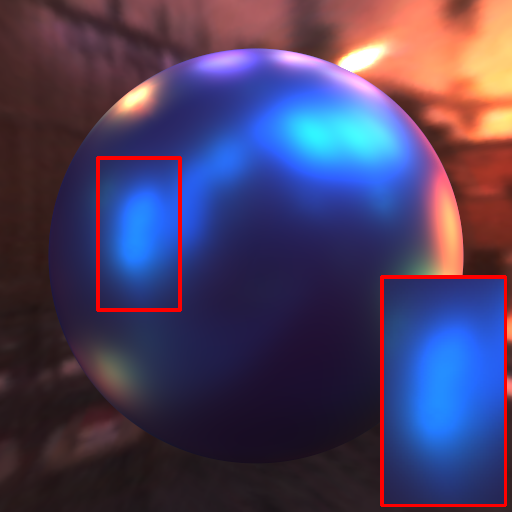} &
        \includegraphics[width=0.135\linewidth]{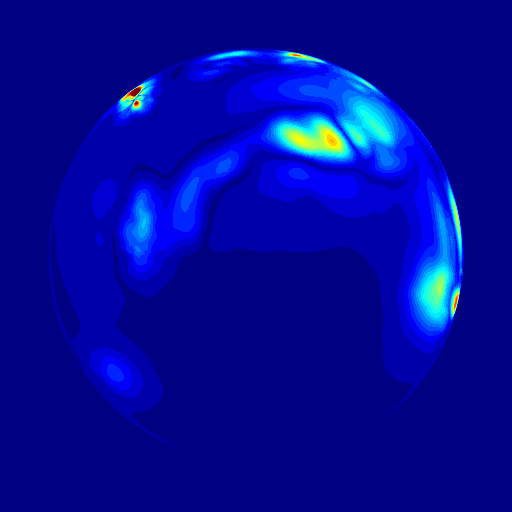} &
        \raisebox{+2\height}{\includegraphics[width=0.13\linewidth]{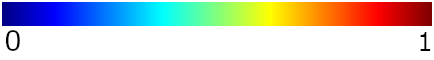}}
        \\ 
        \multicolumn{2}{c}{Nielsen-Trial 4} & \multicolumn{2}{c}{Nielsen-Trial 5} & \multicolumn{2}{c}{FROST-BRDF} &          
        \\ 
    \end{tabular}
    \caption{\label{fig:ring_effect} Rendered images of the reconstructed layered material  \cite{Jakob14}, \textit{vch-silk-blue-rgb}, with $m = 10$ samples. Here we show the robustness of our method in comparison to Nielsen et al. \cite{Nielsen2015}, where vastly different results are obtained on each trial of their proposed method. The absolute error images are placed on the right of each method. Ring artifacts are noticeable in Nielsen-Trial 1 and Nielsen-Trial 2. The error images are multiplied by 10.}
\end{figure*}

\begin{figure}[ht]
\begin{center}
\includegraphics[width=0.85\linewidth]{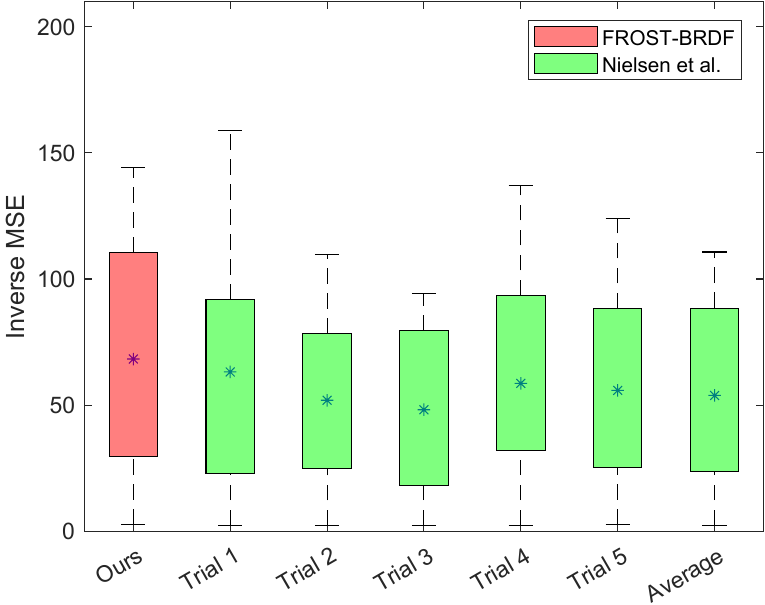}
\end{center}
\caption{\label{fig:our_vs_6dtu} Each box contains 25 percentile (bottom of the box), 75 percentile (head of the box), maximum value (top of the dashed line), minimum value (bottom of the dashed line) and its mean (star mark inside the box). The plots are based on reconstruction error with 40 samples.}
\end{figure}

\begin{figure}[ht]
\begin{center}
\includegraphics[width=1.0\linewidth]{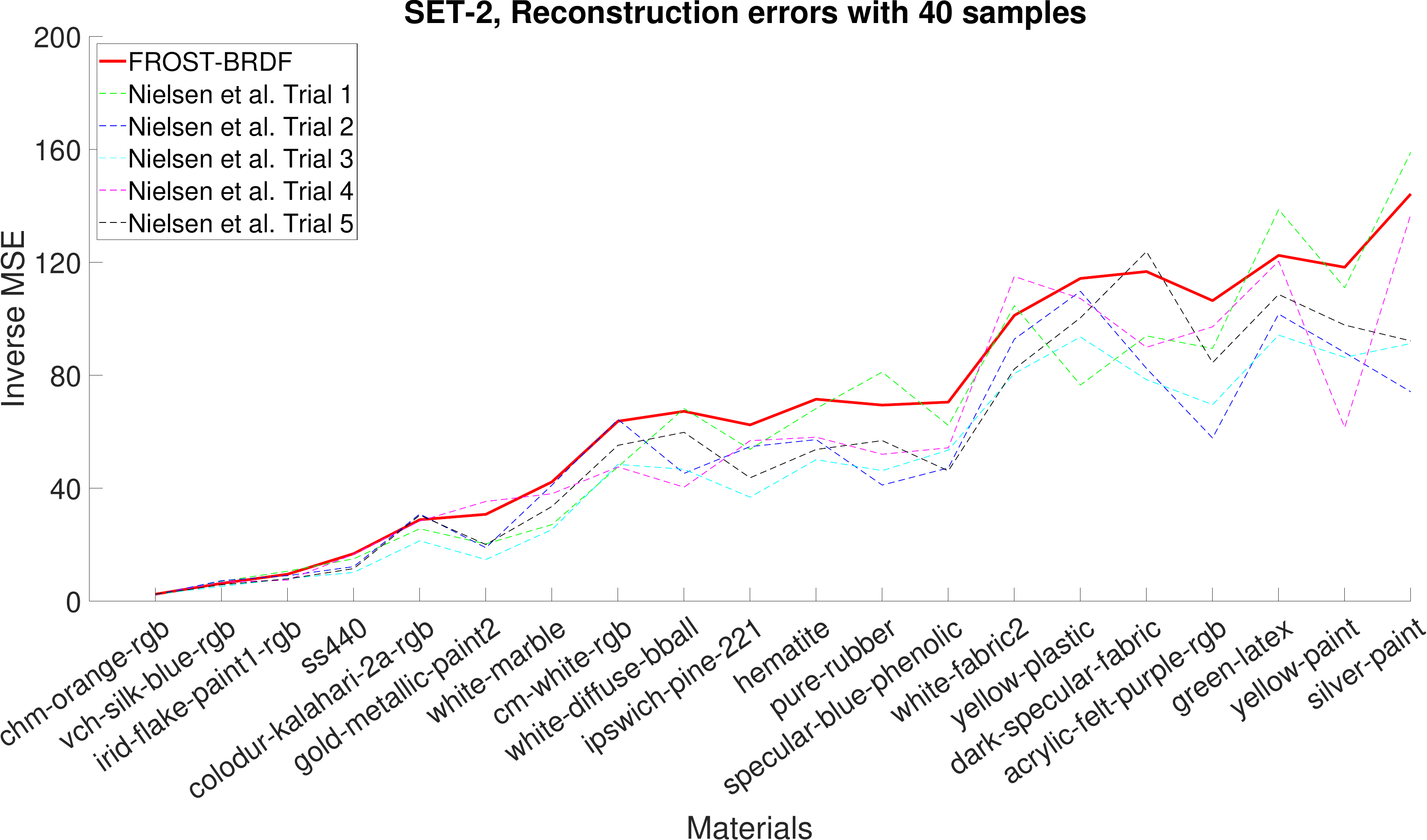}
\end{center}
\caption{\label{fig:our_vs_5dtu_line} Line chart of inverse MSE comparing between our (the red line) and five trials of Nielsen et al. \cite{Nielsen2015} (the dashed lines). The plots are based on reconstruction error with 40 samples.}
\end{figure}

\begin{figure*}[ht] 
    \centering
    \resizebox{0.8\textwidth}{!}{%
    \setlength{\tabcolsep}{0.01cm}
    \setlength\extrarowheight{0.01cm}
    \renewcommand{\arraystretch}{1}
    {
    \begin{tabular}{cccccc}
        \textbf{  } & \textbf{$m=10$} & \textbf{$m=20$} & \textbf{$m=30$} & \textbf{$m=40$} & {Reference} \\ 
        \multirow{2}{*}[-0.2cm]{ \hspace{-0.7cm} \textit{\rot{ specular-blue-phenolic}}} &
        \tabcell{\includegraphics[width=0.147\linewidth,valign=t]{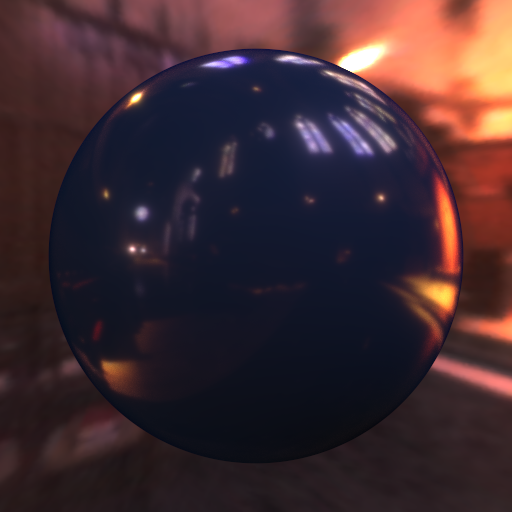}} &
        \tabcell{\includegraphics[width=0.147\linewidth,valign=t]{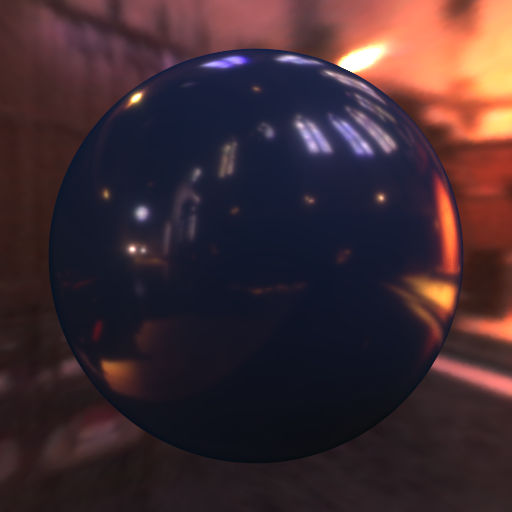}} &
        \tabcell{\includegraphics[width=0.147\linewidth,valign=t]{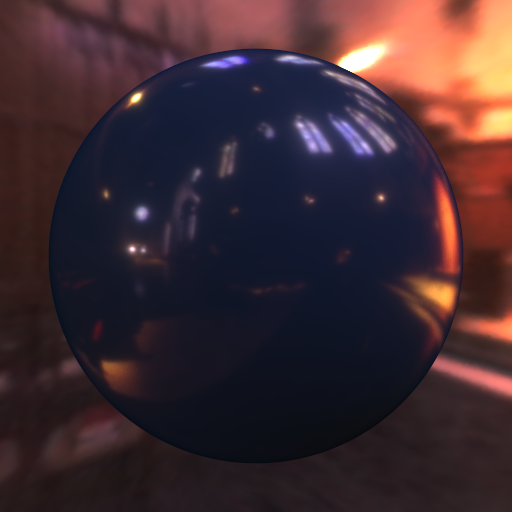}} &
        \tabcell{\includegraphics[width=0.147\linewidth,valign=t]{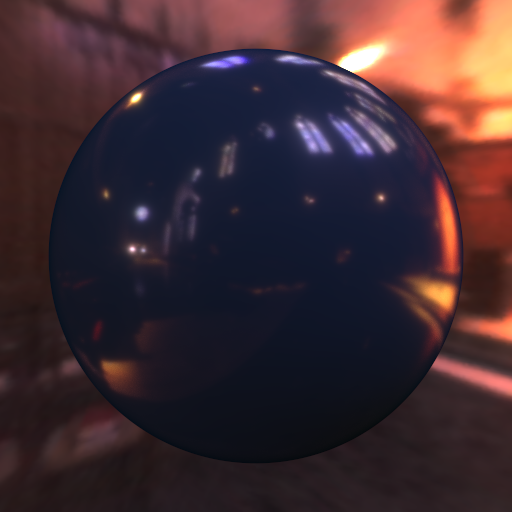}} &
        \tabcell{\includegraphics[width=0.147\linewidth,valign=t]{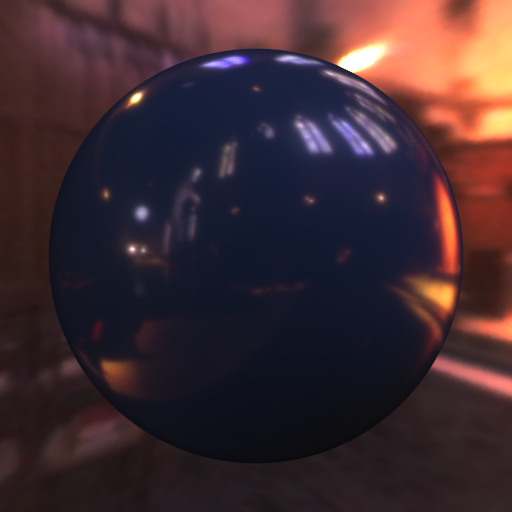}} 
        \\ 
         &
        \tabcell{\includegraphics[width=0.147\linewidth,valign=t]{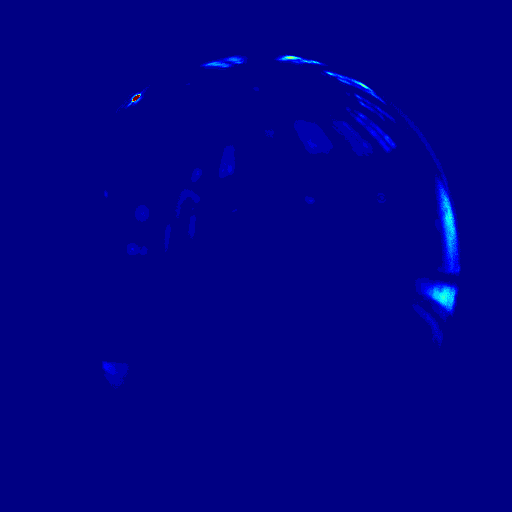}} &
        \tabcell{\includegraphics[width=0.147\linewidth,valign=t]{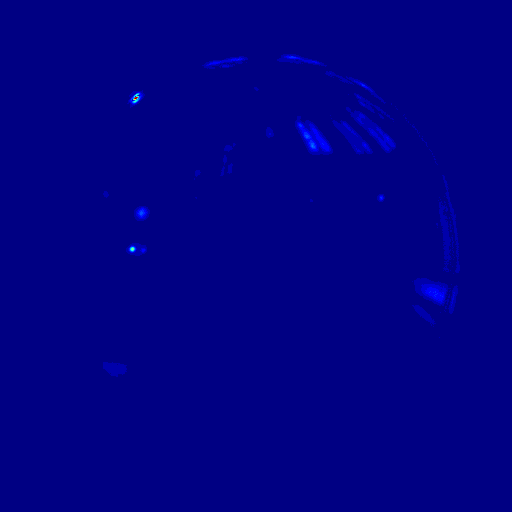}} &
        \tabcell{\includegraphics[width=0.147\linewidth,valign=t]{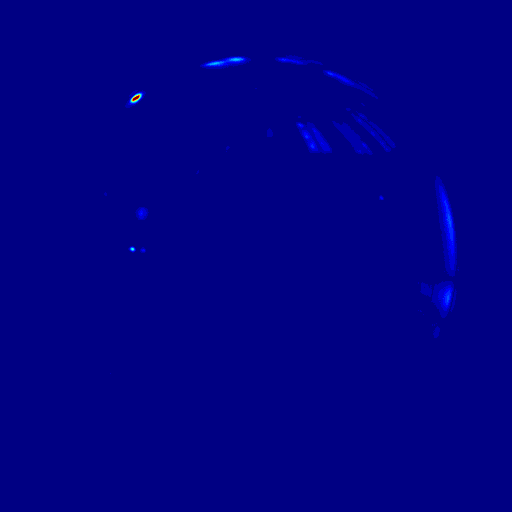}} &
        \tabcell{\includegraphics[width=0.147\linewidth,valign=t]{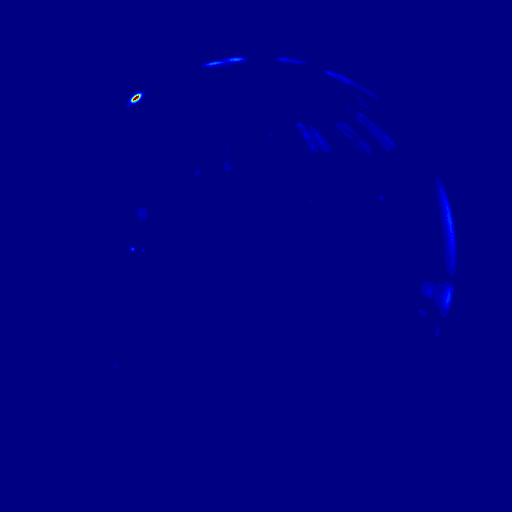}} &
        \tabcell{\includegraphics[width=0.147\linewidth,valign=t]{fig/colormap_jet1.png}}   
        \\  
        \multirow{2}{*}[-0.5cm]{ \hspace{-0.7cm} \textit{\rot{ ipswich-pine-221}}} &
        \tabcell{\includegraphics[width=0.147\linewidth,valign=t]{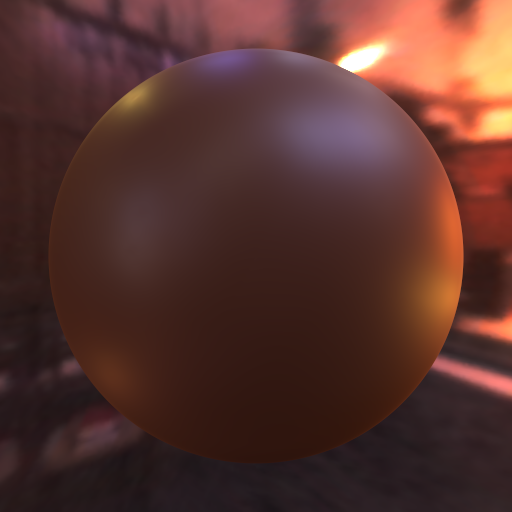}} &
        \tabcell{\includegraphics[width=0.147\linewidth,valign=t]{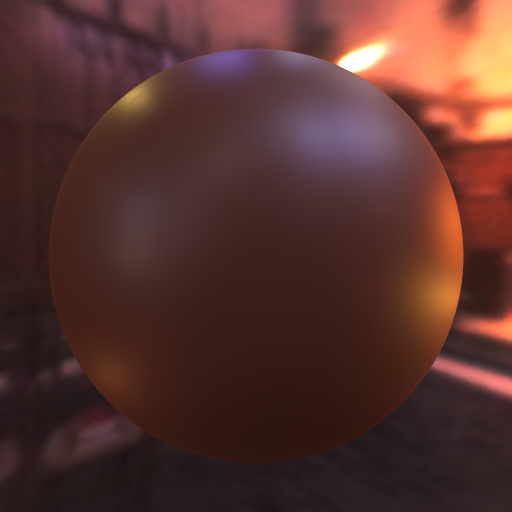}} &
        \tabcell{\includegraphics[width=0.147\linewidth,valign=t]{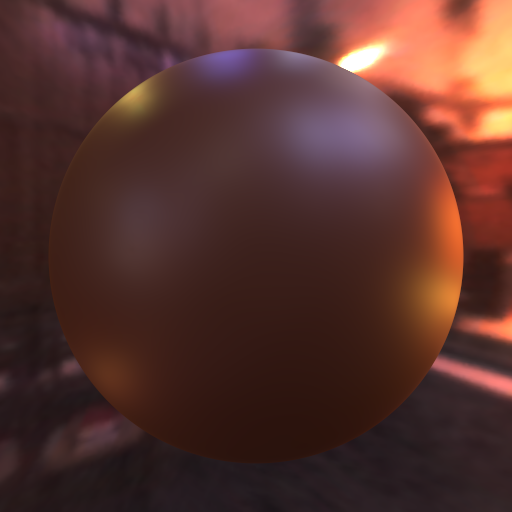}} &
        \tabcell{\includegraphics[width=0.147\linewidth,valign=t]{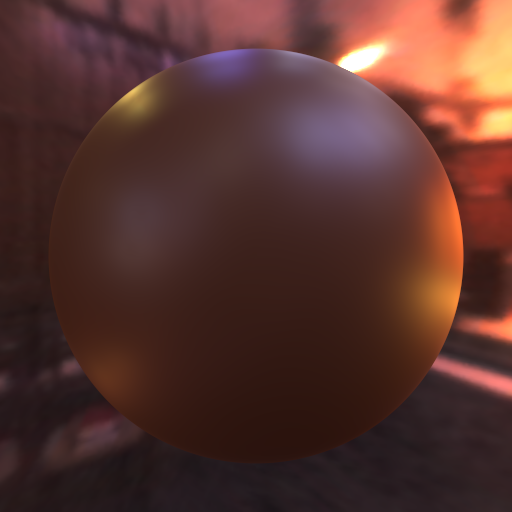}} &
        \tabcell{\includegraphics[width=0.147\linewidth,valign=t]{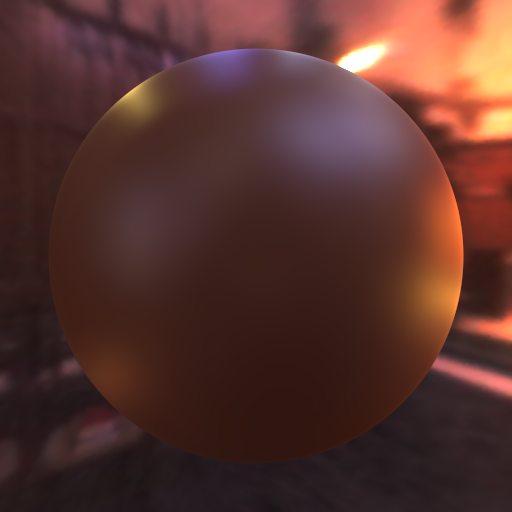}} 
        \\ 
         &
        \tabcell{\includegraphics[width=0.147\linewidth,valign=t]{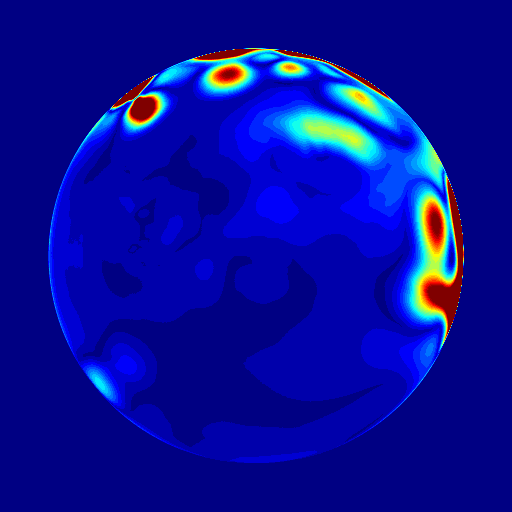}} &
        \tabcell{\includegraphics[width=0.147\linewidth,valign=t]{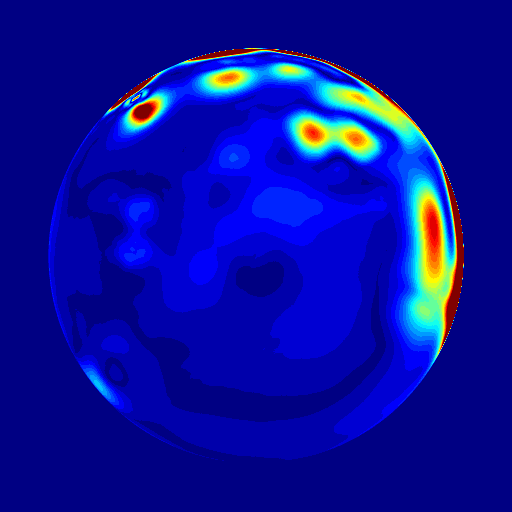}} &
        \tabcell{\includegraphics[width=0.147\linewidth,valign=t]{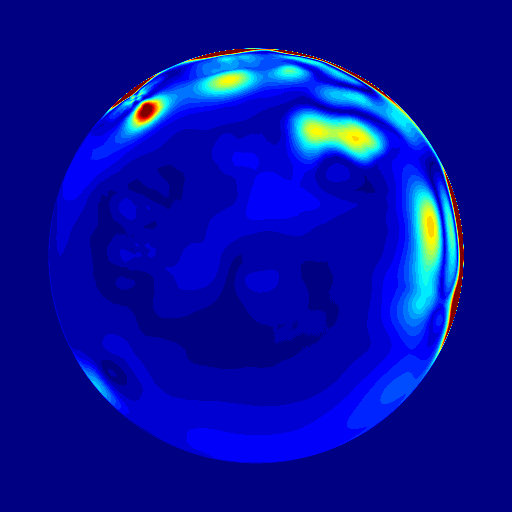}} &
        \tabcell{\includegraphics[width=0.147\linewidth,valign=t]{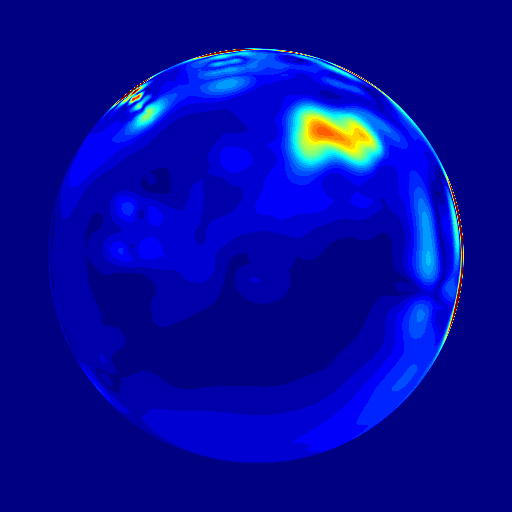}} &
        \tabcell{\includegraphics[width=0.147\linewidth,valign=t]{fig/colormap_jet1.png}}   
        \\ 
    
    \end{tabular}%
    }
    }
    \caption{Two examples of reconstructed BRDFs using FROST-BRDF. We used $m = 10, 20, 30,$ and $40$ samples. The second row, below each rendered scene, shows the corresponding absolute error images normalized across four rendered images. The error images are multiplied by 10 to facilitate comparisons.}
    \label{fig:our_examples}
\end{figure*}

\begin{figure*}[ht]
    \centering
    \resizebox{1.00\linewidth}{!}{%
    \setlength{\tabcolsep}{0.002cm}
    \setlength\extrarowheight{-3pt}
    \small
    \begin{tabular}{cccccccccc}
        \hspace{14pt}  & 
        \multicolumn{2}{c}{\textit{hematite}} &
        \multicolumn{2}{c}{\textit{green-latex}} & 
        \multicolumn{2}{c}{\textit{gold-metallic-paint2}} & 
        \multicolumn{2}{c}{\textit{cm-white-rgb}} &
        
        \\ 
        \raisebox{0.05\height }{\rotatebox[origin=l]{90}{\textbf{Nielsen et al.}}} &
        \includegraphics[width=0.115\linewidth]{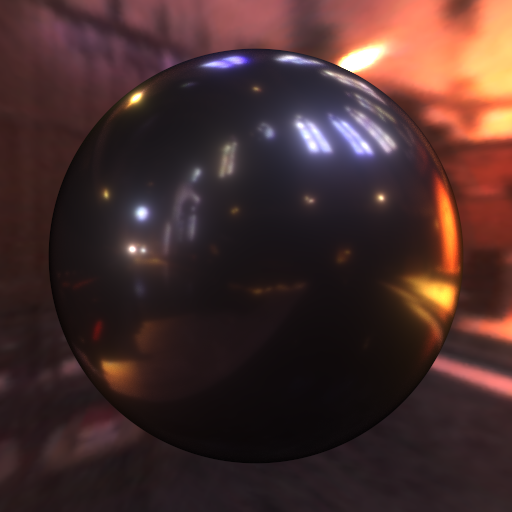} &
        \includegraphics[width=0.115\linewidth]{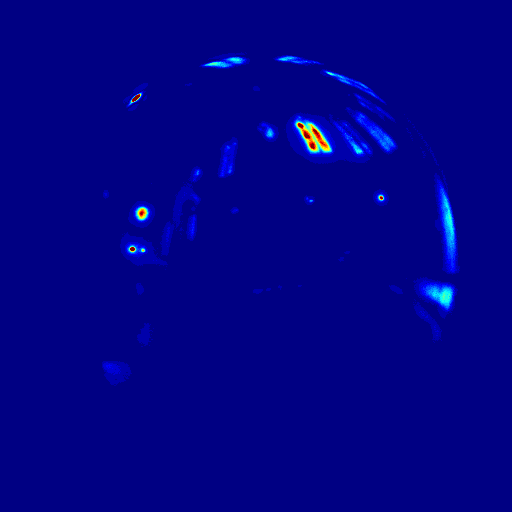} &
        \includegraphics[width=0.115\linewidth]{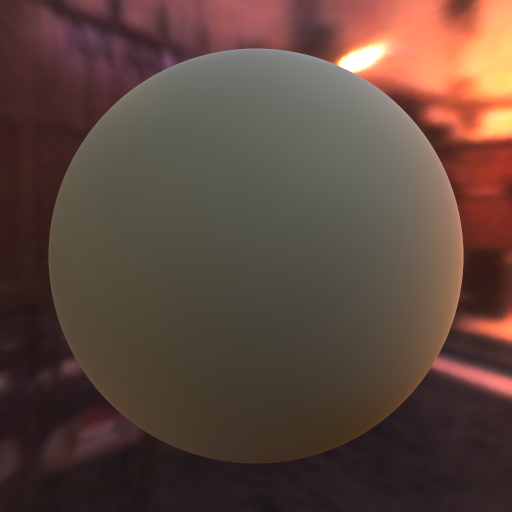} &
        \includegraphics[width=0.115\linewidth]{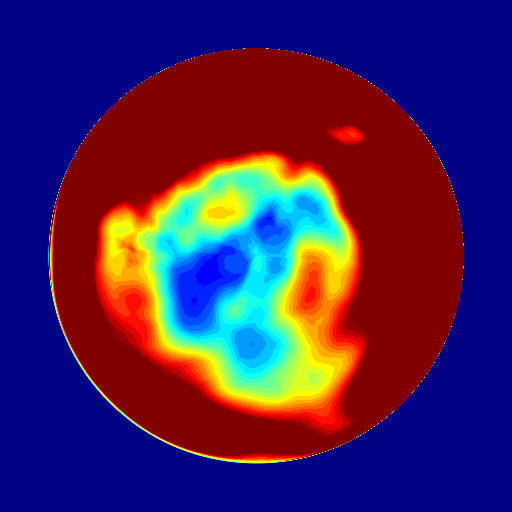} &
        \includegraphics[width=0.115\linewidth]{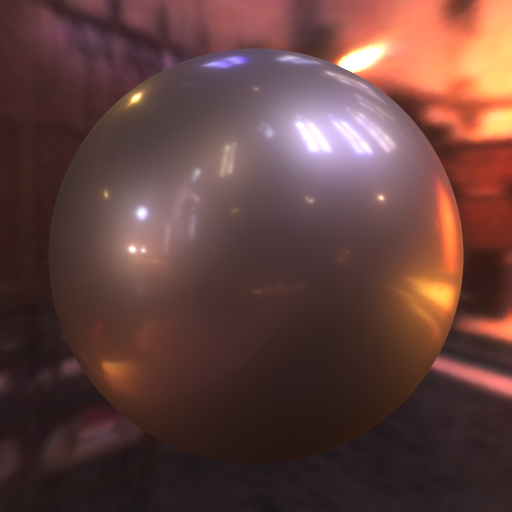} &
        \includegraphics[width=0.115\linewidth]{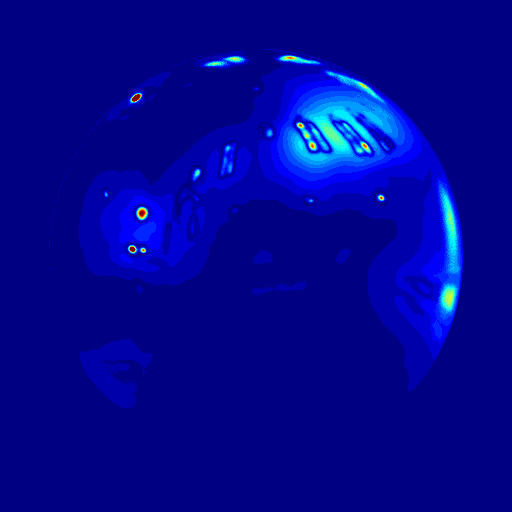} &
        \includegraphics[width=0.115\linewidth]{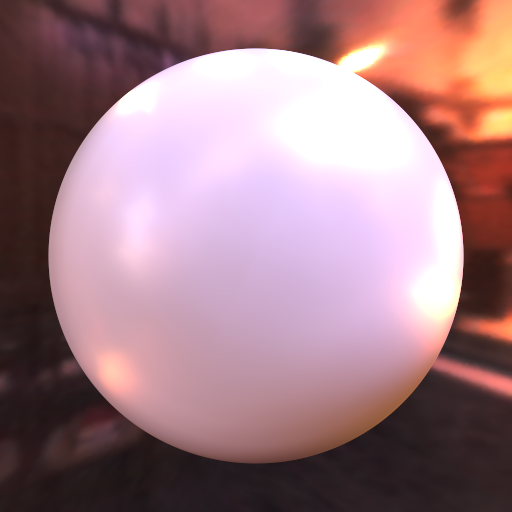} &
        \includegraphics[width=0.115\linewidth]{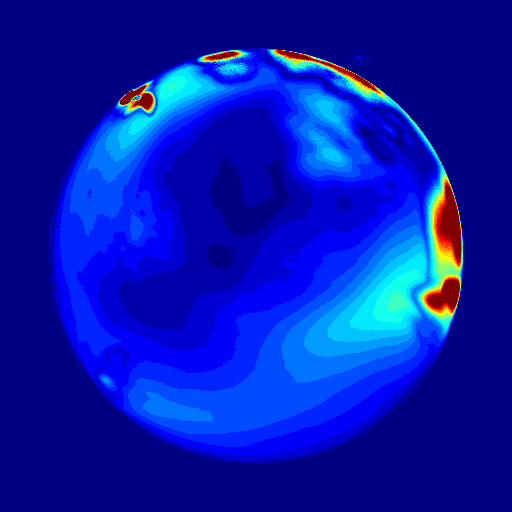} &
        \hspace{0.01cm}\includegraphics[width=0.0147\linewidth]{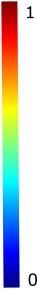}
        \\
         & \small{35.40dB}
         &
         & \small{44.98dB}
         &
         &\small{34.45dB}
         &
         &\small{25.75dB}
         &
         &
        \\
        \raisebox{0.01\height }{\rotatebox[origin=l]{90}{\textbf{FROST-BRDF}}} &
        \includegraphics[width=0.115\linewidth]{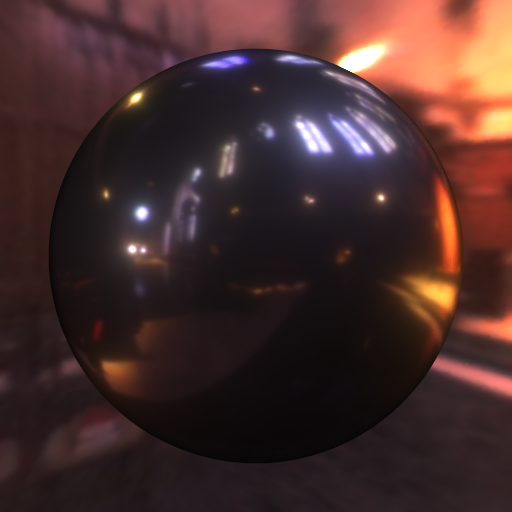} &
        \includegraphics[width=0.115\linewidth]{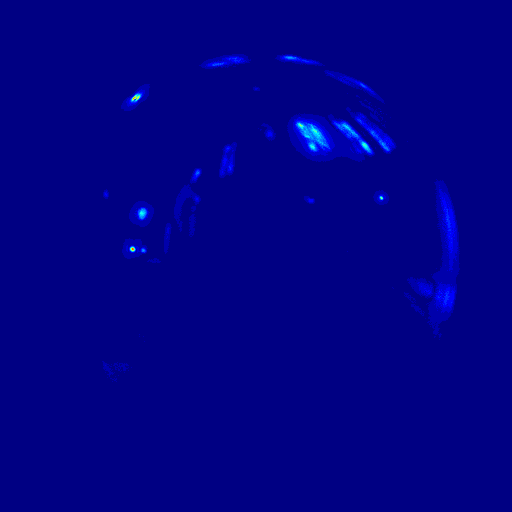} &
        \includegraphics[width=0.115\linewidth]{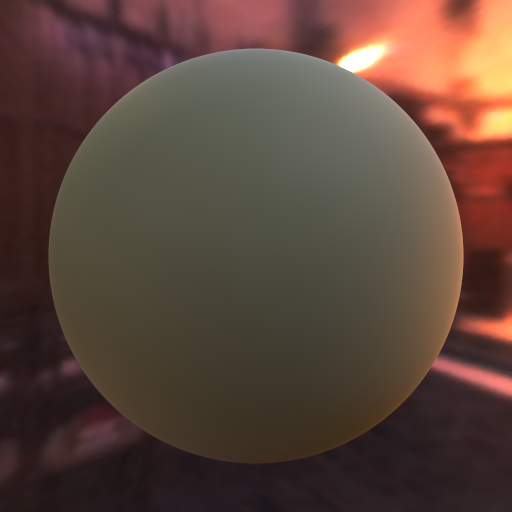} &
        \includegraphics[width=0.115\linewidth]{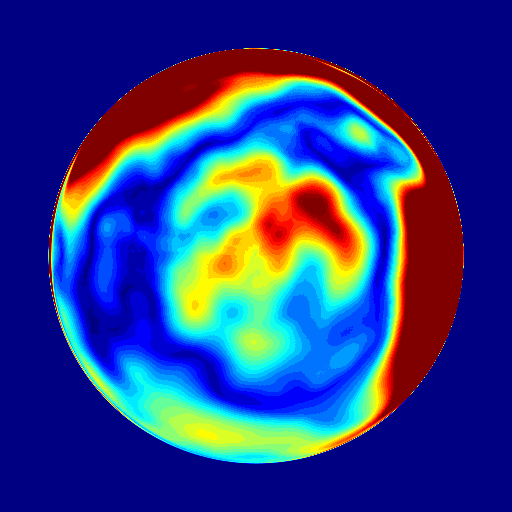} &
        \includegraphics[width=0.115\linewidth]{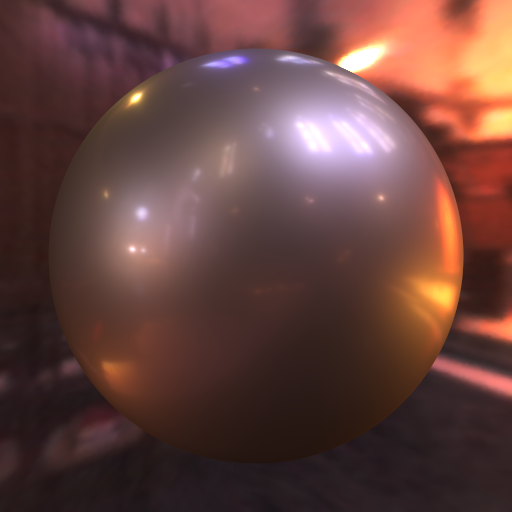} &
        \includegraphics[width=0.115\linewidth]{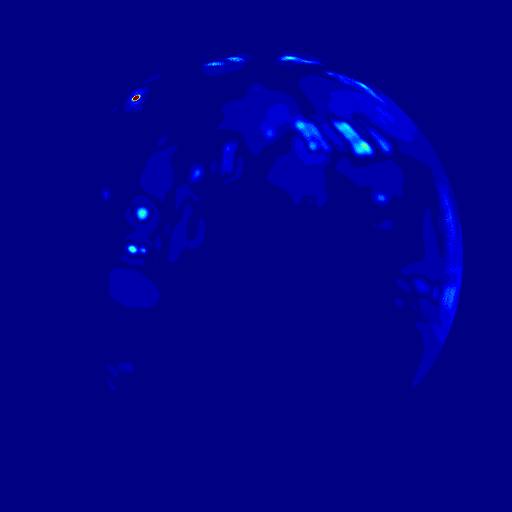} &
        \includegraphics[width=0.115\linewidth]{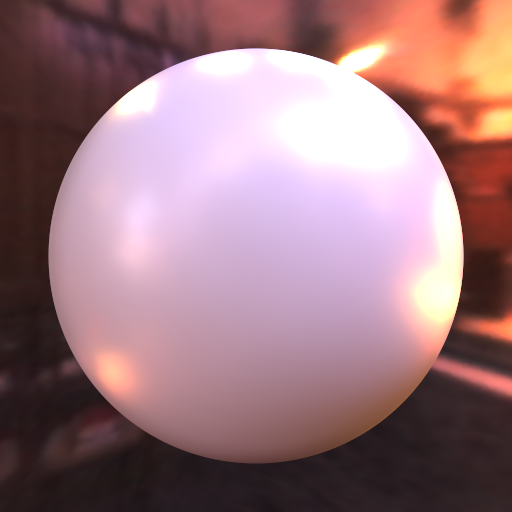} &
        \includegraphics[width=0.115\linewidth]{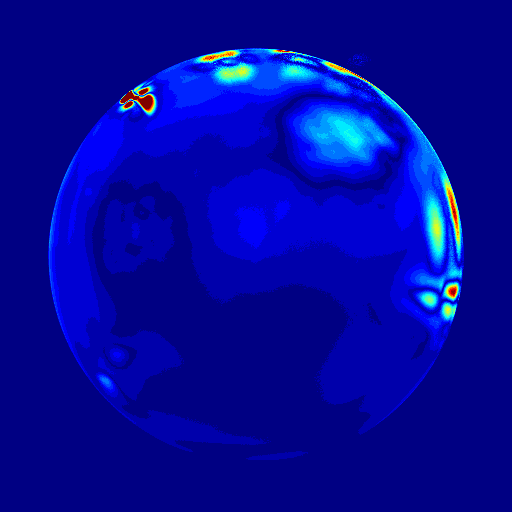} &
        \hspace{0.01cm}\includegraphics[width=0.0147\linewidth]{fig/colormap_jet_vert.png}
        \\
         & \small{44.81dB}
         &
         & \small{51.62dB}
         &
         & \small{41.76dB}
         &
         &  \small{32.16dB}
         &
         &
        \\

        \raisebox{0.25\height }{\rotatebox[origin=l]{90}{\textbf{Reference}}} &
        \includegraphics[width=0.115\linewidth]{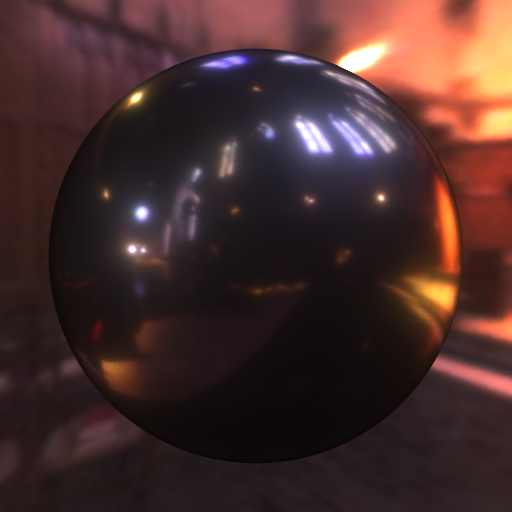} &
         &
        \includegraphics[width=0.115\linewidth]{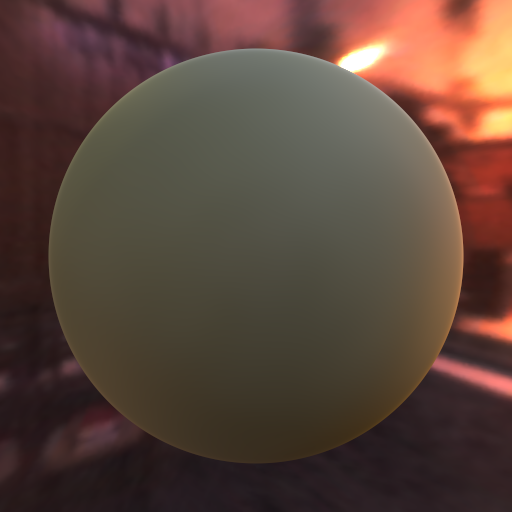} &
         &
        \includegraphics[width=0.115\linewidth]{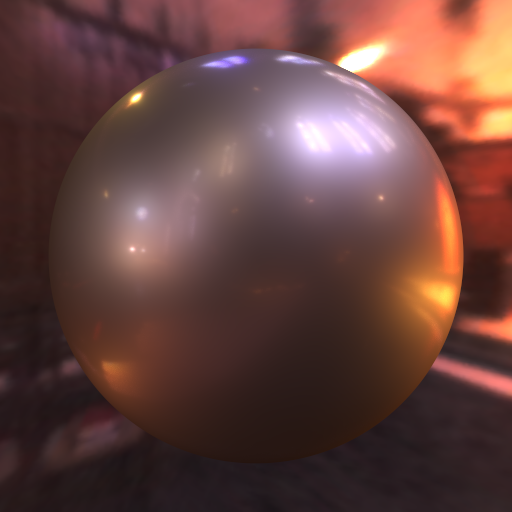} &
         &
        \includegraphics[width=0.115\linewidth]{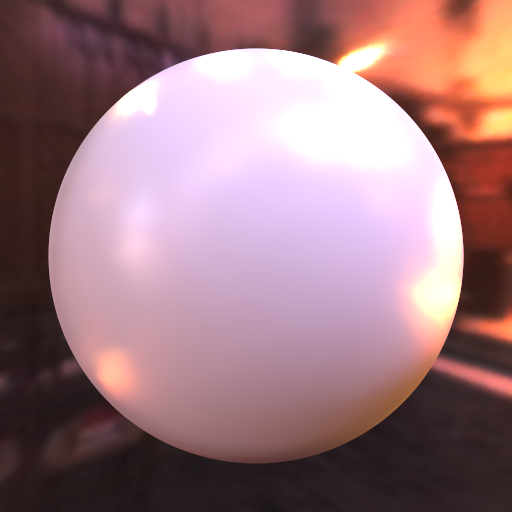} &
         &
        
        \\ 
    \end{tabular}
    }
    
    \caption{\label{fig:our_vs_dtu_examples} Four examples of reconstructed BRDFs with $m = 20$. The first row shows the reconstructions from Nielsen et al. \cite{Nielsen2015} and the second row shows our reconstructions using FROST-BRDF. The absolute error images are multiplied by 10 to facilitate comparisons. We also report rendering SNR for each method.}
\end{figure*}

\begin{figure*}[ht]
    \centering
    \resizebox{1.00\linewidth}{!}{%
    \setlength{\tabcolsep}{0.002cm}
    \setlength\extrarowheight{-3pt}
    \small
    \begin{tabular}{cccccccccc}
        \hspace{14pt}  & 
        \multicolumn{2}{c}{\textit{glossy-red-paper}} &
        \multicolumn{2}{c}{\textit{green-cloth}} & 
        \multicolumn{2}{c}{\textit{notebook}} & 
        \multicolumn{2}{c}{\textit{yellow-paper}} &
        
        \\ 
        \raisebox{0.7\height }{\rotatebox[origin=l]{90}{\textbf{m=20}}} &
        \includegraphics[width=0.115\linewidth]{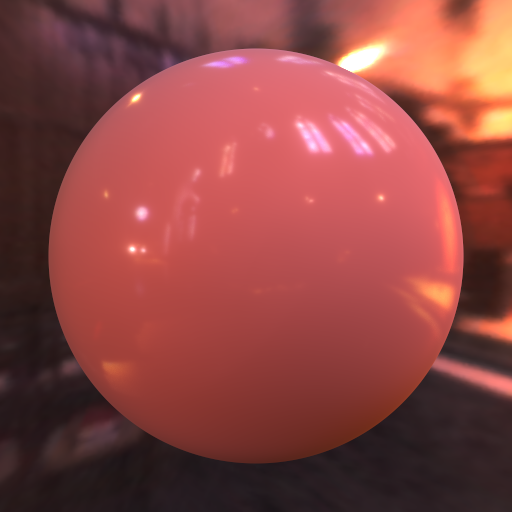} &
        \includegraphics[width=0.115\linewidth]{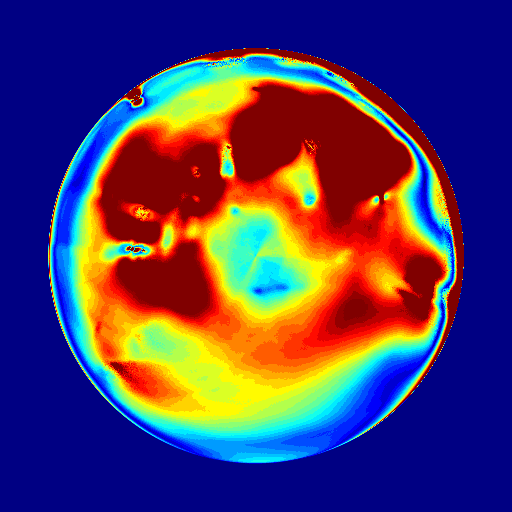} &
        \includegraphics[width=0.115\linewidth]{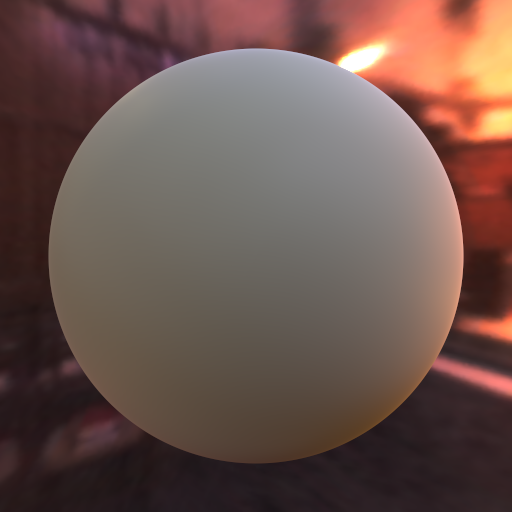} &
        \includegraphics[width=0.115\linewidth]{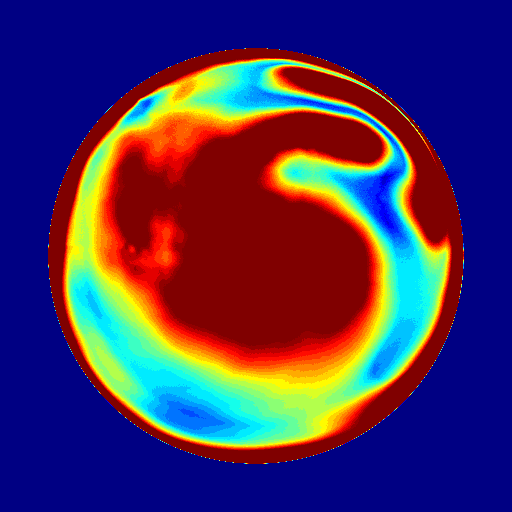} &
        \includegraphics[width=0.115\linewidth]{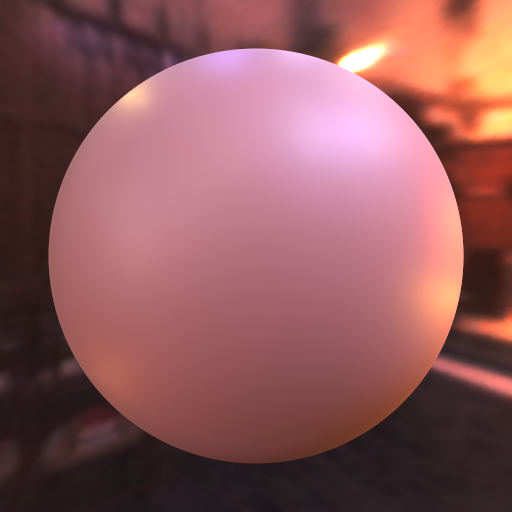} &
        \includegraphics[width=0.115\linewidth]{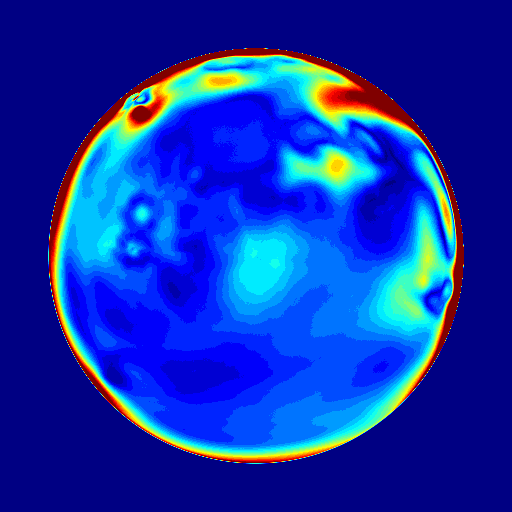} &
        \includegraphics[width=0.115\linewidth]{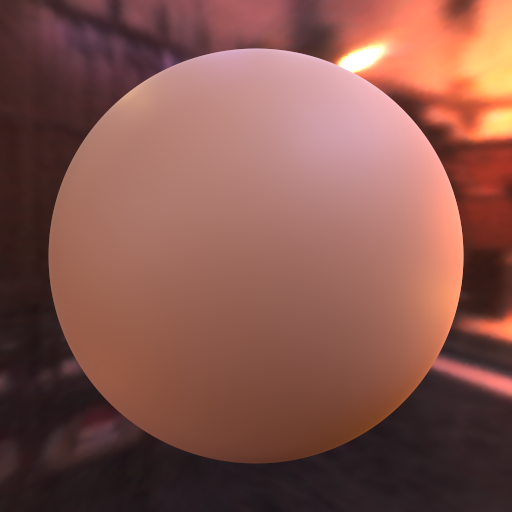} &
        \includegraphics[width=0.115\linewidth]{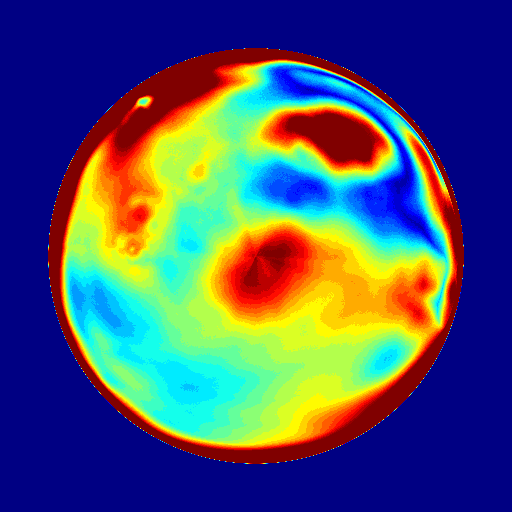} &
        \hspace{0.01cm}\includegraphics[width=0.0147\linewidth]{fig/colormap_jet_vert.png}
        \\
         & \small{43.50dB}
         &
         & \small{50.84dB}
         &
         &\small{42.16dB}
         &
         &\small{46.12dB}
         &
         &
        \\
        \raisebox{0.7\height }{\rotatebox[origin=l]{90}{\textbf{m=40}}} &
        \includegraphics[width=0.115\linewidth]{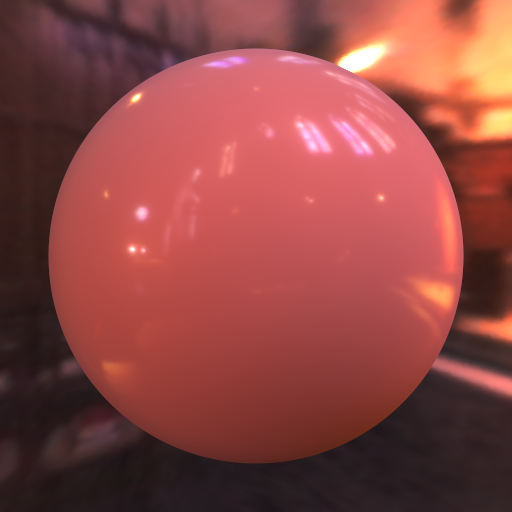} &
        \includegraphics[width=0.115\linewidth]{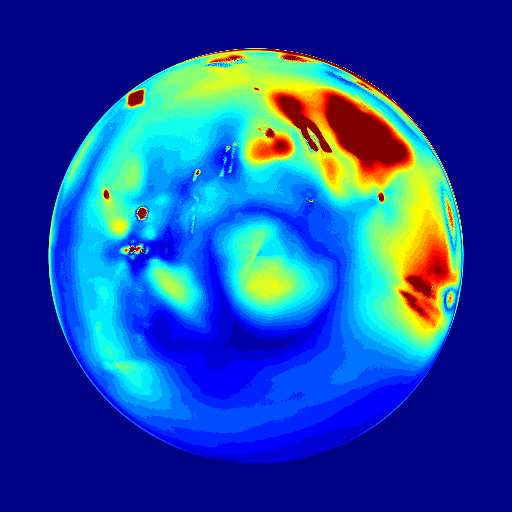} &
        \includegraphics[width=0.115\linewidth]{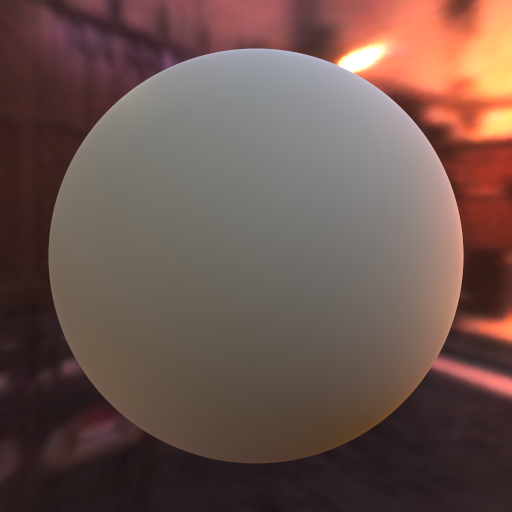} &
        \includegraphics[width=0.115\linewidth]{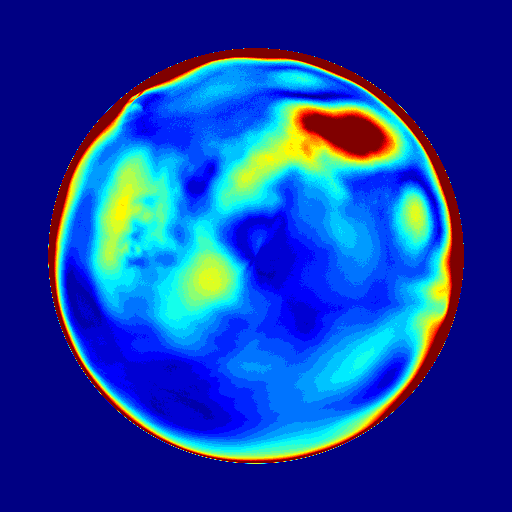} &
        \includegraphics[width=0.115\linewidth]{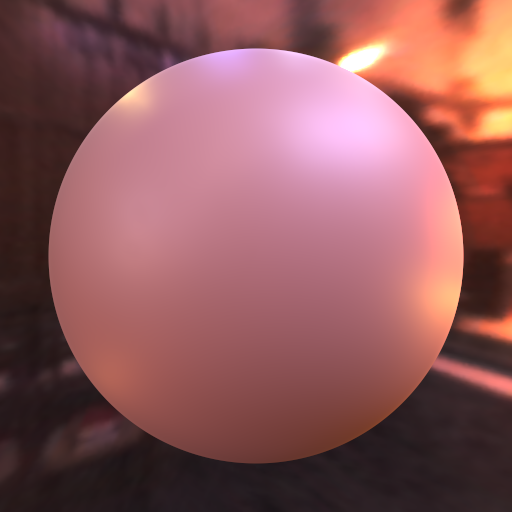} &
        \includegraphics[width=0.115\linewidth]{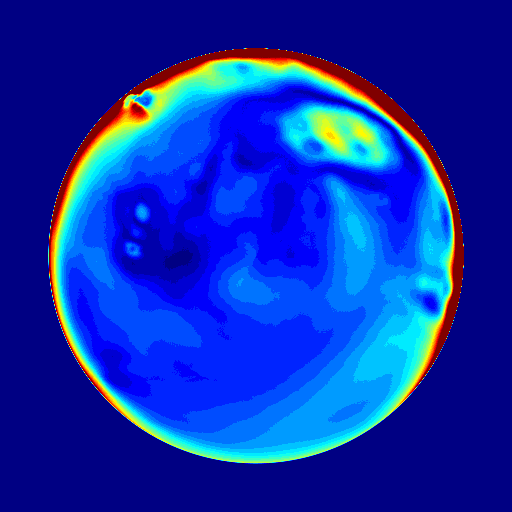} &
        \includegraphics[width=0.115\linewidth]{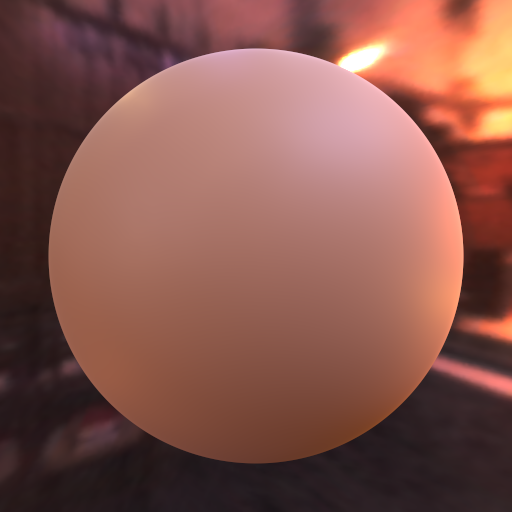} &
        \includegraphics[width=0.115\linewidth]{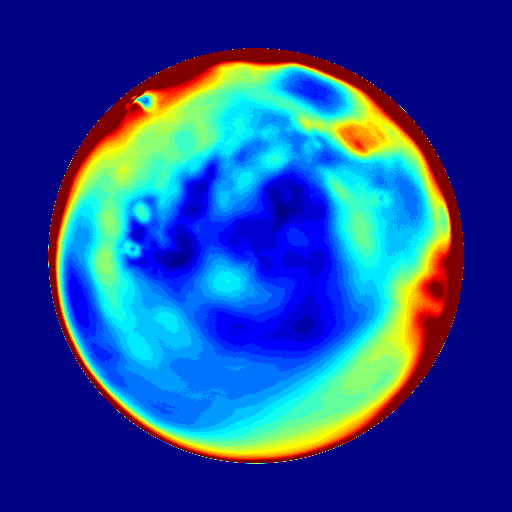} &
        \hspace{0.01cm}\includegraphics[width=0.0147\linewidth]{fig/colormap_jet_vert.png}
        \\
         & \small{48.93dB}
         &
         & \small{54.38dB}
         &
         & \small{44.02dB}
         &
         &  \small{47.73dB}
         &
         &
        \\

        \raisebox{0.2\height }{\rotatebox[origin=l]{90}{\textbf{Reference}}} &
        \includegraphics[width=0.115\linewidth]{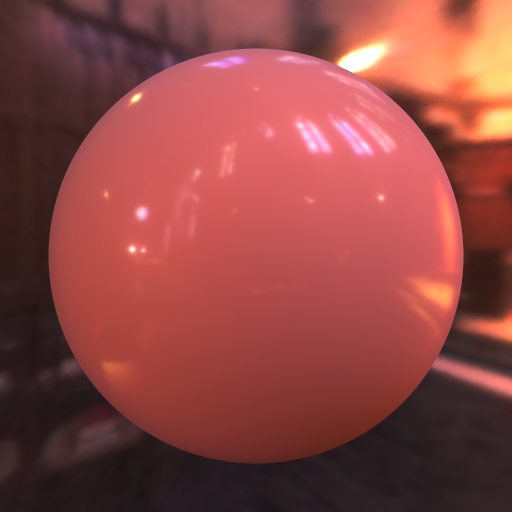} &
         &
        \includegraphics[width=0.115\linewidth]{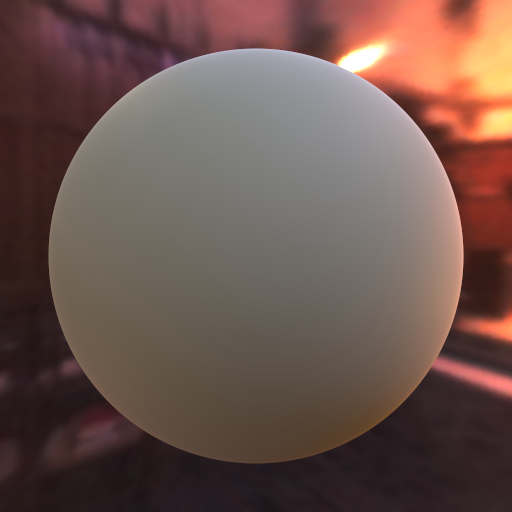} &
         &
        \includegraphics[width=0.115\linewidth]{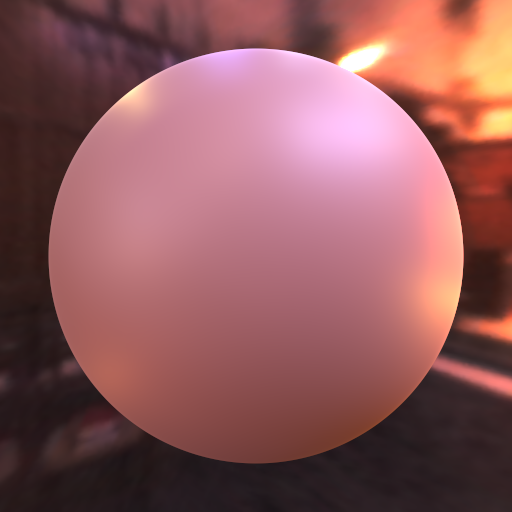} &
         &
        \includegraphics[width=0.115\linewidth]{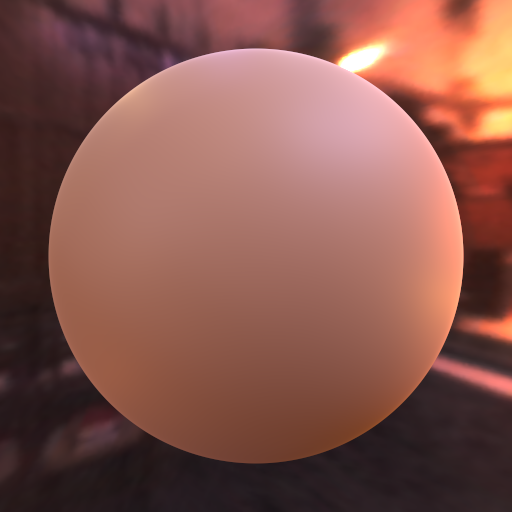} &
         &
        
        \\ 
    \end{tabular}
    }
    
    \caption{\label{fig:our_on_dtu_dataset} Four examples of reconstructed BRDFs using FROST with $m=20$ and $m=40$ performed over the DTU dataset \cite{Nielsen2015}. No materials from this dataset were included in our training set, which demonstrates the robustness of FROST-BRDF.}
\end{figure*}

\begin{figure}[ht]
    \begin{center}
    \includegraphics[width=0.95\linewidth]{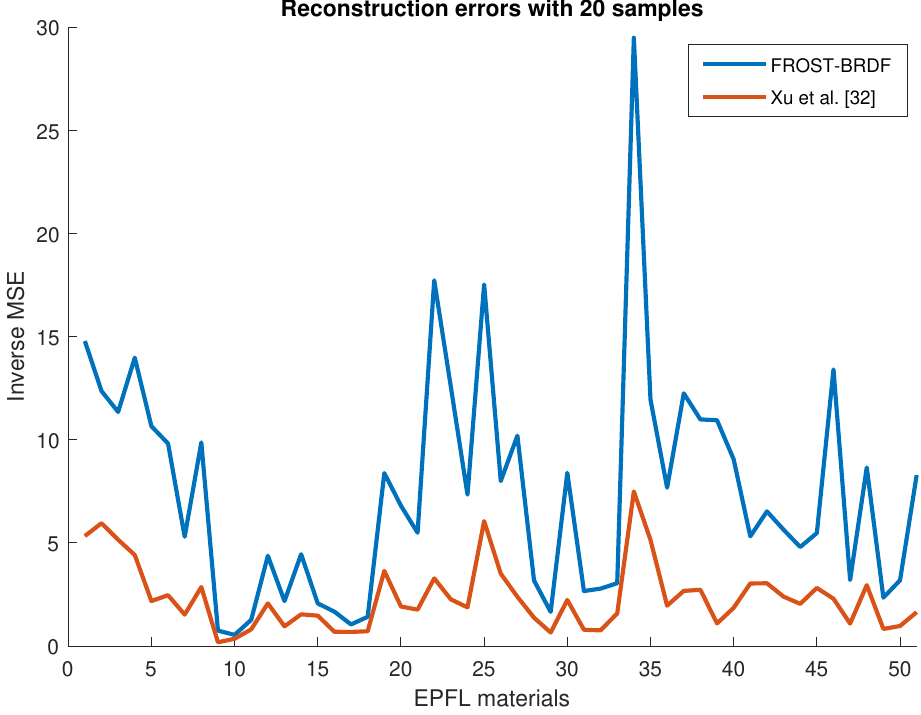}
    \end{center}
    \caption{Numerical comparisons between FROST-BRDF and the method of Xu et al. \cite{Xu2016:sampling} with respect to 51 BRDFs from the RGL-EPFL data set \cite{Dupuy:EPFL}. FROST-BRDF achieves a higher reconstruction quality for all BRDFs within this data set. Both methods were tested using 20 optimal BRDF sampling directions. For the method of Xu et al. \cite{Xu2016:sampling}, we used the optimal directions provided by the authors. \label{fig:xu}}
\end{figure}

\begin{figure}[ht]
    \centering
    \resizebox{1.00\linewidth}{!}{%
    \setlength{\tabcolsep}{0.002cm}
    \setlength\extrarowheight{-3pt}
    \small
    \begin{tabular}{cccc}
            \raisebox{0.1\height }{\rotatebox[origin=l]{90}{\textbf{Xu et al. \cite{Xu2016:sampling}}}} \hspace{1pt} &
            \includegraphics[width=0.3\linewidth]{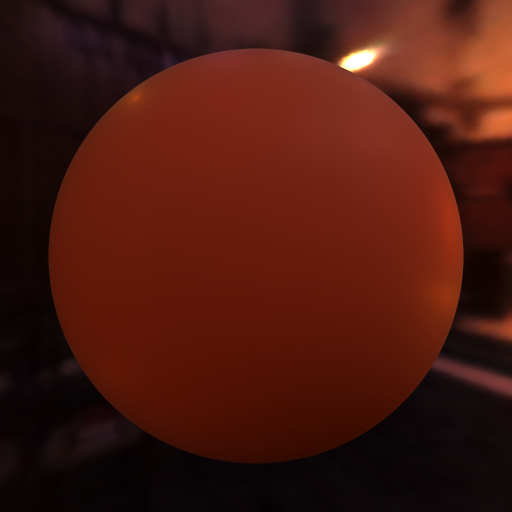} &
			\includegraphics[width=0.3\linewidth]{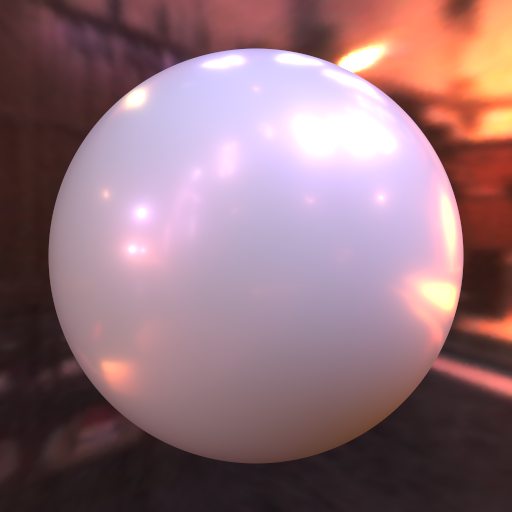}  &
			\includegraphics[width=0.3\linewidth]{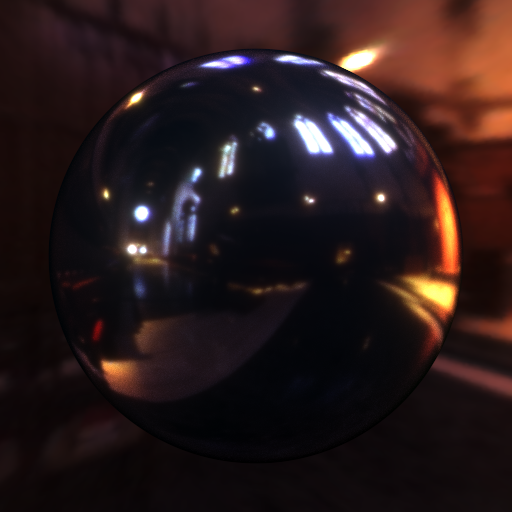} \\
            & $23.10$dB & $26.96$dB & $24.66$dB \\
            \raisebox{0.8\height }{\rotatebox[origin=l]{90}{\textbf{Error}}} \hspace{1pt} &
            \includegraphics[width=0.3\linewidth]{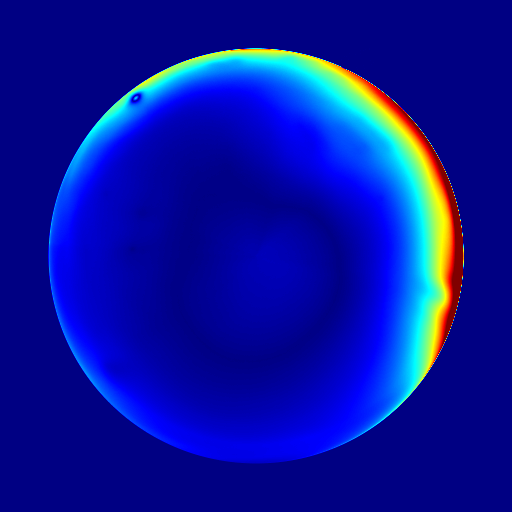} &
			\includegraphics[width=0.3\linewidth]{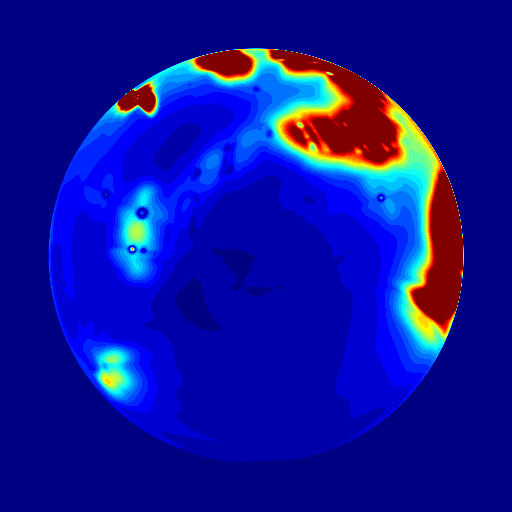} &
			\includegraphics[width=0.3\linewidth]{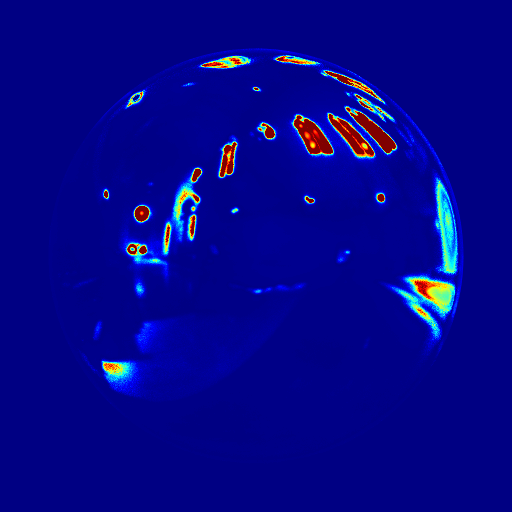} \\
            \raisebox{0.1\height }{\rotatebox[origin=l]{90}{\textbf{FROST-BRDF}}} \hspace{1pt} &
			\includegraphics[width=0.3\linewidth]{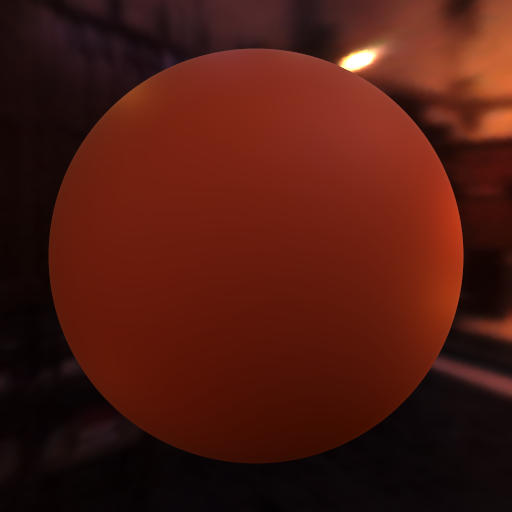} &
			\includegraphics[width=0.3\linewidth]{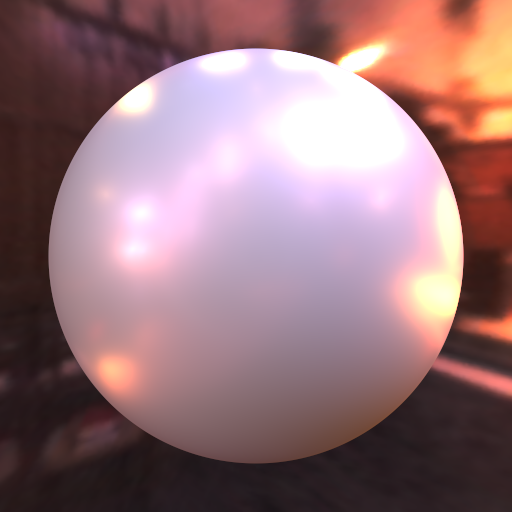}  &
			\includegraphics[width=0.3\linewidth]{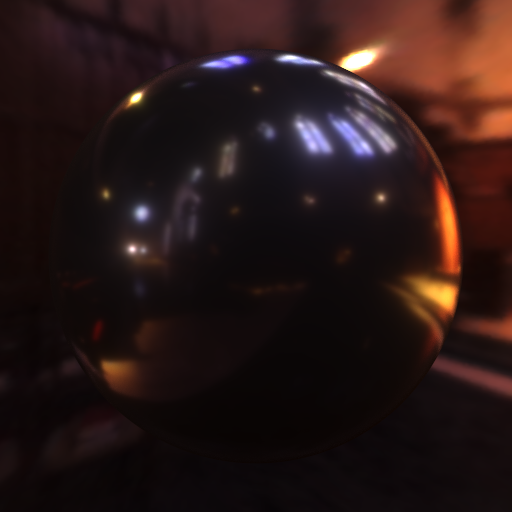} \\
            &  $27.26$dB  & $31.35$dB & $32.78$dB \\
            \raisebox{0.8\height }{\rotatebox[origin=l]{90}{\textbf{Error}}} \hspace{1pt} &
			\includegraphics[width=0.3\linewidth]{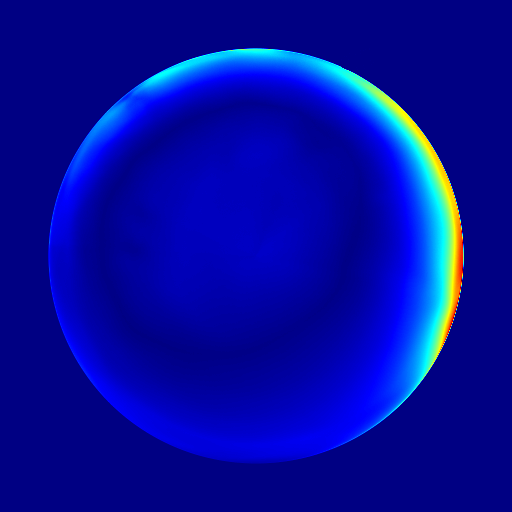} &
			\includegraphics[width=0.3\linewidth]{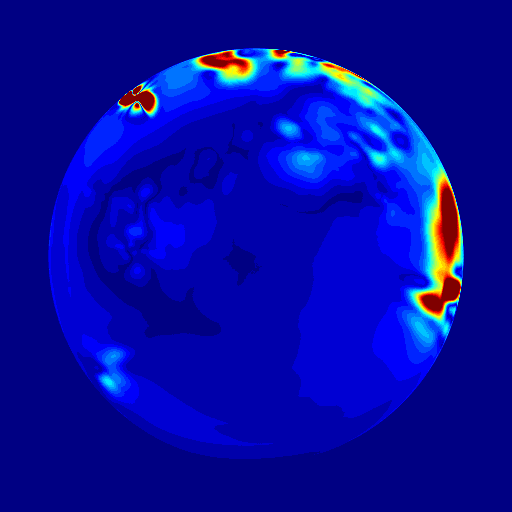} &
			\includegraphics[width=0.3\linewidth]{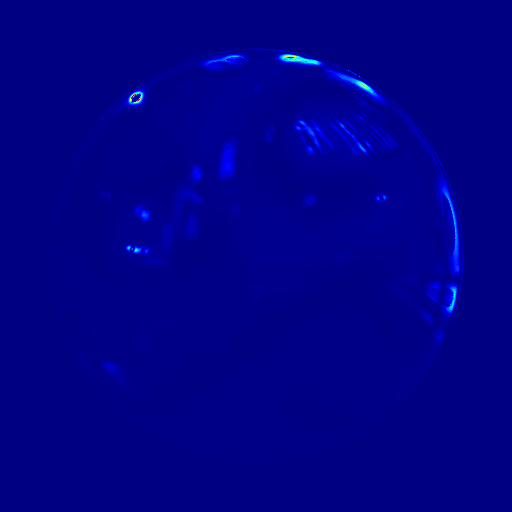} \\
            \raisebox{0.3\height }{\rotatebox[origin=l]{90}{\textbf{Reference}}} \hspace{1pt} &
            \includegraphics[width=0.3\linewidth]{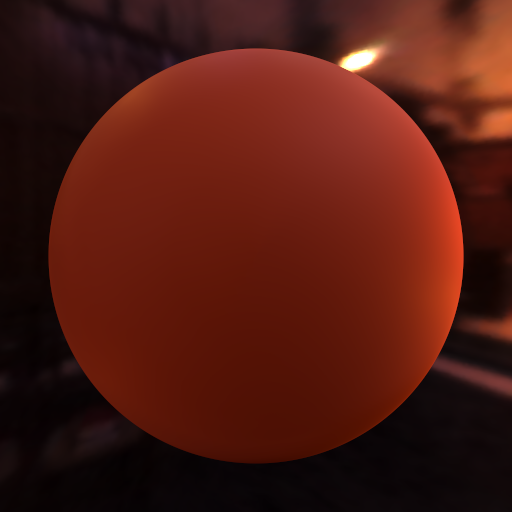} &
			\includegraphics[width=0.3\linewidth]{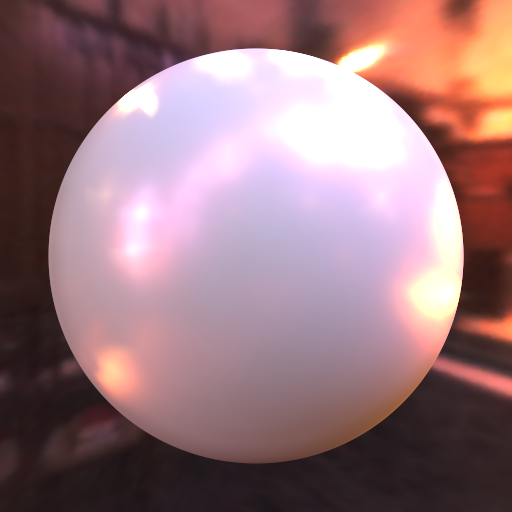}  &
			\includegraphics[width=0.3\linewidth]{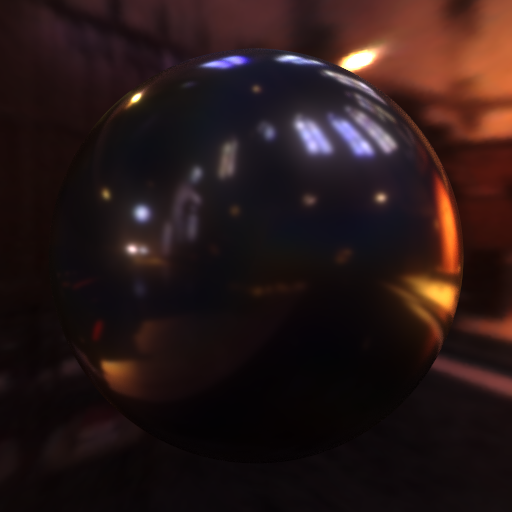} \\
    \end{tabular}
    }
    \caption{\label{fig:xu-visual} Visual comparison results for FROST-BRDF and the method of Xu et al. \cite{Xu2016:sampling} using $20$ sample directions. From left to right, the materials are \emph{acrylic-felt-orange-rgb}, \emph{satin-rosaline-rgb}, and \emph{irid-flake-paint1-rgb}, all from the RGL-EPFL data set. We used the MERL data set for the training of both methods. Rendering SNR is reported below each image.}
\end{figure}

\section{Results and Discussions}   \label{sec:result}

We evaluate FROST-BRDF through a series of comparisons with the method of Nielsen et al. \cite{Nielsen2015} that was proposed for recovering a full BRDF from a small number of measurements. Our evaluations were performed on two publicly available datasets, MERL \cite{Matusik2003:MERL}, and RGL-EPFL \cite{Dupuy:EPFL}, as they provide measured isotropic BRDFs. The MERL and RGL-EPFL datasets contain 100 and 51 materials, respectively. 

We randomly split the dataset into 131 materials for training, and 20 materials for testing as our base experiment setting. Out of 131 materials included in the training set, 86 and 45 materials were collected from the MERL and RGL-EPFL datasets, respectively. We evaluated our results with a 10-fold cross-validation each with an individual random training - test set split across the datasets. In this setup, the training data matrix $\mat{T}$, introduced in Section \ref{sec:method:formulation}, has $n = 90 \times 90 \times 180$ rows, and $t = 131 \times 3 = 393$ columns, since each BRDF has three color channels. Note that the total number of valid BRDF values is $1,111,432$. Therefore, in practice, the size of the matrices $\mat{T}$ and $\mat{D}$ are $1,111,432\times393$, which amounts to $1.63$GB of storage, assuming a $32$-bit floating point storage. 

To evaluate FROST-BRDF, and the methods we compare to, we report results with $m = 5$ to $m=60$, with an incremental step of $5$. As mentioned in Section \ref{sec:impl:recon}, since we set $m=k$ for a fair comparison with \cite{Nielsen2015}, we do not need to tune the number of principal components, $k$. The effect of $k$ and $m$ on the reconstruction quality is shown in Figure \ref{fig:p-m}.

Our method was implemented in MATLAB and the source code for FROST-BRDF accompanies this paper\footnote{\url{https://github.com/emiandji/frost_brdf}}. After the reconstruction of a full BRDF using FROST from optimal samples, we render the result using PBRT \cite{Pharr:2010:PBR}, followed by tone mapping the image with a Gamma function, where $\gamma = 2.0$. The HDR environment map used here is \textit{Grace Cathedral}. The absolute error images were normalized with respect to the maximum absolute pixel error for our method and those we compare to.

Since the optimization method of Nielsen et al. \cite{Nielsen2015} is based on a random row selection of the dictionary $\mat{D}$ at each iteration, we performed five trials for each training - test set split to test variations between each run. Since our method is deterministic, the results reported here are obtained with one trial. In what follows, we discuss our evaluations regarding the performance and accuracy of both methods. The supplementary materials accompanying this manuscript contain a thorough and detailed presentation of experiments for a larger set of materials.

\subsection{Average quantitative visual quality}
Figure \ref{fig:dtu_vs_our} shows the BRDF reconstruction error of our method with respect to the baseline method \cite{Nielsen2015}. For both methods, $131$ materials were included in the training set. We calculated Mean Squared Error (MSE) over the transformed BRDF domain, i.e., using the \textit{log-relative mapping} function. The plot shows the inverse of MSE on the vertical axis, and the number of samples on the horizontal axis. The higher the inverse MSE value, the better the reconstruction quality over the BRDF domain. We observe that from $10$ samples, our method (blue plot) performs significantly better than the baseline method (orange plot). We also evaluated our method with only $80$ materials included in training sets (red plot), making the training set size smaller than that of the baseline method by a ratio of 0.4. This result indicates that our approach performs with similar accuracy as the baseline method when a much smaller number of materials are included in the training set. Note that the results reported in Figure \ref{fig:dtu_vs_our} are the mean performance of FROST-BRDF and the baseline method using a $10$-fold cross-validation. We performed an extensive number of experiments to evaluate the robustness of our algorithm for all 10 sets, and our method outperforms the state-of-the-art in all such sets. The detailed plots of different cross-validation sets are shown in the supplementary material. The results reported in the remainder of this Section are based on the second cross-validation set, which we call SET-2 in the supplementary document.

\subsection{Consistency}
In Table \ref{tab:our_dtu_5samples}, we show an example of optimal sampling directions obtained by FROST, together with 5 trials of the baseline method. The baseline method, due to random initializations at each iteration, generated diverse sampling directions in each trial. A visual comparison of the rendering results obtained by the method of Nielsen et al. for the five aforementioned trials, in comparison to our method, is shown in Figure \ref{fig:ring_effect}, where we use $m = 10$ samples for both methods. The BRDF used here is a layered material, specifically \textit{vch-silk-blue-rgb}, that represents a challenging scenario for acquisition. The ringing artifacts in the highlights are evident on Nielsen-Trial 1, 2, 3, and 4, which were also demonstrated in \cite{Nielsen2015}. In contrast, our method does not show such artifacts, which indicates the robustness of FROST-BRDF with a low number of samples, as well as the consistency of the result without the need for performing multiple trials. 
The inconsistency of the results obtained from the baseline method using 5 trials is further exemplified in Figure \ref{fig:our_vs_6dtu}. All the statistical quantities vary significantly as the experiment changes from Trial 1 to Trial 5. However, our method, shown in red, robustly outperforms all trials of Nielsen et al. \cite{Nielsen2015}. The corresponding detailed line plot of this experiment, with various materials, is shown in Figure \ref{fig:our_vs_5dtu_line}. The line plot indicates that each test material reconstructed with our method (the red line), in the vast majority of cases, is superior in reconstruction quality compared to 5 trials of the baseline method. It can also be observed from Figure \ref{fig:our_vs_5dtu_line}, that even the best performing trial of the baseline method is considerably inferior, on average, in terms of reconstruction quality when compared to FROST-BRDF. Since Nielsen et al.'s method \cite{Nielsen2015} is based on random initializations at each iteration, in rare occasions and for some materials, their method may achieve a slightly higher quality. This is expected, however, we argue that only the average MSE or SNR can be relied on when comparing a deterministic and a random optimization technique.

\subsection{Visual quality and number of samples}
Figure \ref{fig:our_examples} demonstrates two reconstructed materials obtained by FROST-BRDF, namely \textit{specular-blue-phenolic} and \textit{ipswich-pine-221}, where we use $m=10$, 20, 30, and 40 samples. As the rendering result suggests, increasing the number of samples can decrease the errors obtained from the reconstruction, which is expected. The reconstructions of both materials indicate that the error is, for the most part, evenly distributed, except for the highlights at grazing angles which are difficult to reconstruct correctly. This is due to the measurement inaccuracy, i.e. noise and outliers, of the grazing angles for the datasets we use here during training. However, increasing the number of samples can attenuate such reconstruction errors, which shows the robustness of our method to noise and outliers. Note that the absolute error images are multiplied by a factor of 10 for display purposes. 


\subsection{Visual quality w.r.t. Nielsen et al. \cite{Nielsen2015}}
Figure \ref{fig:our_vs_dtu_examples}, shows four examples of the reconstructed BRDFs with $m = 20$ samples. The results from Nielsen et al. \cite{Nielsen2015} appear in the first row, and those from our FROST-BRDF method are in the second row. The glossy materials, such as \textit{specular-blue-phenolic} and \textit{gold-metallic-paint2}, and the diffuse materials, such as \textit{green-latex}, can be reconstructed faithfully with our FROST-BRDF compared to the baseline method. Even in the layered material, \textit{cm-white-rgb}, that is known to be challenging to model, our method achieves a rendering SNR of $32.16$dB. 

\begin{table*}[ht]
    \centering
    \caption{Average optimization time for finding optimal sample directions, measured in seconds, for FROST-BRDF and Nielsen et al. \cite{Nielsen2015}.}
    \small
    \begin{tabular}{|l|c|c|c|c|c|c|c|} 
        \hline
        $m$ & 5 & 10 &  20 &  30 &  40 &  50 &  60 \\ 
        \hline
        FROST-BRDF & 0.14 & 0.47 &  1.86 &  4.21 &  7.49 &  12.05 &  17.67 \\ 
        \hline
        Nielsen et al. \cite{Nielsen2015} & 76.01 & 248.67 &  772.74 &  1600.46 &  2772.74 &  4357.08 &  6366.88 \\
        \hline
    \end{tabular}
    \label{tab:timer}
\end{table*}

\subsection{Additional results}
In Figure \ref{fig:our_on_dtu_dataset}, we report results of FROST-BRDF on the DTU dataset provided by Nielsen et al. \cite{Nielsen2015}. To further demonstrate the robustness of our method, we utilize the same dictionary used for the previously reported results; i.e. none of the materials from the DTU dataset were included in the training set. Even for a glossy material such as \textit{glossy-red-paper}, FROST-BRDF produces high quality reconstructions. 

We present our results in comparison to the method of Xu et al. \cite{Xu2016:sampling} in Figure \ref{fig:xu}, where we report MSE for $51$ materials from the RGL-EPFL data set. For all the materials, FROST-BRDF achieves a lower reconstruction error. The results reported for the method of Xu et al. \cite{Xu2016:sampling} utilize the optimal sample directions provided by the authors that were obtained using a dictionary trained on the MERL data set. We used the same data set for training FROST-BRDF. Even though the reconstruction step is similar for FROST-BRDF and the method of Xu et al., and that the dictionaries are identical, we observe that FROST-BRDF produces significantly higher reconstruction quality. We associate this to the superiority of the sample directions computed by FROST. Figure \ref{fig:xu-visual} compares the visual quality of FROST-BRDF and the method of Xu et al. for three materials from the RGL-EPFL data set, where we also report rendering SNR. We have included six more materials in the supplementary document accompanying this paper.

Finally, in Table \ref{tab:timer}, we present our optimization time compared with Nielsen et al. \cite{Nielsen2015}, using the python source code provided by the authors. We can see that FROST-BRDF finds the optimal sampling directions $370$ times faster, on average, than the method of Nielsen et al. \cite{Nielsen2015}. Both methods were run on a machine with an Intel Xeon W-2123, clocked at 3.60GHz.

\section{Limitations and Future Work} \label{sec:limitation}

In our proposed framework, we find only one sensing matrix for all types of materials in a training set. However, due to the diversity of material properties, it is more reasonable to find separate sampling patterns for each BRDF family. Indeed, this requires clustering of the training set, ideally based on a metric that takes the FROST approximation error into account. The main issue that arises in this case, is the need to determine the material cluster prior to acquisition. A possible solution is to design a neural network for learning the basic properties of the BRDF from a single image to help identify the suitable class of sensing directions given a new material \cite{Henzler2021}. 

It is well-known that the error obtained from compressed sensing is lower-bounded by the error of the sparse representation of the signal we would like to acquire. Therefore, the quality of the sparse representation, hence the dictionary, is of utmost importance. In this paper, to simplify the exposition of our method and to facilitate comparisons with previous work, we employ the most simple signal representation using PCA. However, previous studies have shown that sparse representation using over-complete \cite{marwah2013compressive} or dictionary ensembles \cite{ehsan-tog2019} produce vastly superior results. As a future work direction, it would be interesting to use a dictionary ensemble similar to \cite{ehsan-tog2019} for improving the sampling directions, and consequently the reconstruction quality. Learning-based dictionaries that promote sparsity have been shown to be theoretically \cite{candes-cs-intro} and empirically \cite{ehsan-tog2019,ksvd} superior in reducing the required number of samples.

The results reported in this paper use measured BRDF datasets which contain a moderate amount of noise, hence demonstrating the robustness of FROST-BRDF to noise. However, for future work, we would like to perform a systematic noise analysis to quantify the tolerance of our approach to a range of noise types with different variances and magnitudes. Moreover, the accuracy of our method is dependent on the error metric employed by our sampling and reconstruction algorithms, see e.g. equations \eqref{eq:cs-recon} and \eqref{eq:frost-l2}, where we utilize $\ell_2$ norm for quality assessment, i.e. the fitting error. Previous studies \cite{lavoue:hal-03128383,Tongbuasirilai22} have shown that minimizing the $\ell_2$ norm of linear BRDFs for model fitting does not necessarily lead to good rendering quality. Designing a suitable error metric based on the human visual system would be another interesting future research direction. 


FROST solves an omnipresent problem in science and engineering, i.e. efficient sub-sampling and reconstruction of discrete signals. 
For graphics and imaging applications, FROST can be employed for acquisition of several data modalities such as light fields, light field videos, Bidirectional Texture Functions (BTF), and multidimensional medical and scientific data. Given low computational cost of FROST, it can also be utilized in applications where the sub-sampling operator needs to be updated frequently, e.g. when additional training data is available, or for coded-aperture light field photography \cite{marwah2013compressive,saghi-lfv}. Another interesting venue for future work is the utilization of FROST for designing efficient light-stage capturing systems \cite{lightstage-debevec}, or improving the acquisition speed of existing systems.  


\vspace{-2pt}
\section{Conclusions }   \label{sec:conclusion}
This paper introduced a novel formulation of the BRDF acquisition problem using compressed sensing, as well as a new sampling technique, FROST, for computing a small set of optimal sample locations given a training set. Based on the BRDF samples obtained using FROST, we demonstrated high-quality reconstructions of full BRDFs. We compared FROST-BRDF to the state-of-the-art method proposed by Nielsen et al. \cite{Nielsen2015}, showing that with an equal number of samples, FROST-BRDF achieves significantly lower reconstruction error, as well as higher rendering quality. The 10-fold cross-validation experiments show that our reconstruction results perform consistently in all statistical quantities such as 75 percentile, max and mean values. Additionally, our method is shown to be robust since it produces the optimal samples deterministically, unlike previous work where each run produces vastly different results. Moreover, our FROST-BRDF algorithm is, on average, $370$ times faster than that of \cite{Nielsen2015}. We also demonstrate that since our method is built on the strong mathematical foundation of compressed sensing, we can derive theoretical guarantees. Given the simplicity and generality of FROST, we believe that it can be utilized in a variety of applications that require optimal sampling and reconstruction.


\section{Acknowledgement}\label{sec:acknowledgement}
This project has received funding from the European Union’s Horizon 2020 research and innovation programme under the Marie Skłodowska-Curie grant agreement No. 956585 (PRIME).

\bibliographystyle{IEEEtran}
\bibliography{main.bib}



\begin{IEEEbiography}[{\includegraphics[width=1in,height=1.25in,clip,keepaspectratio]{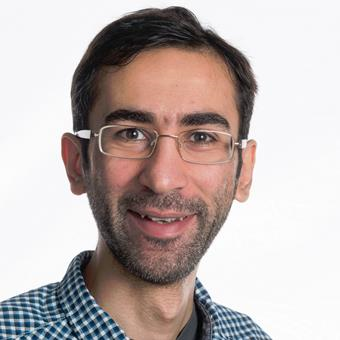}}]{Ehsan Miandji}
received the M.S. and Ph.D. degrees in computer graphics and image processing from Link\"{o}ping University, Sweden, in 2012 and 2018, respectively. In 2020, he pursued postdoctoral research at Inria, Rennes, France. He is currently an assistant professor of computer graphics at the Department of Science and Technology, Linköping University. His research interests include compressive light field photography, compression and compressed sensing of visual data, appearance modeling and acquisition, and photorealistic image synthesis. 
\end{IEEEbiography}

\begin{IEEEbiography}[{\includegraphics[width=1in,height=1.25in,clip,keepaspectratio]{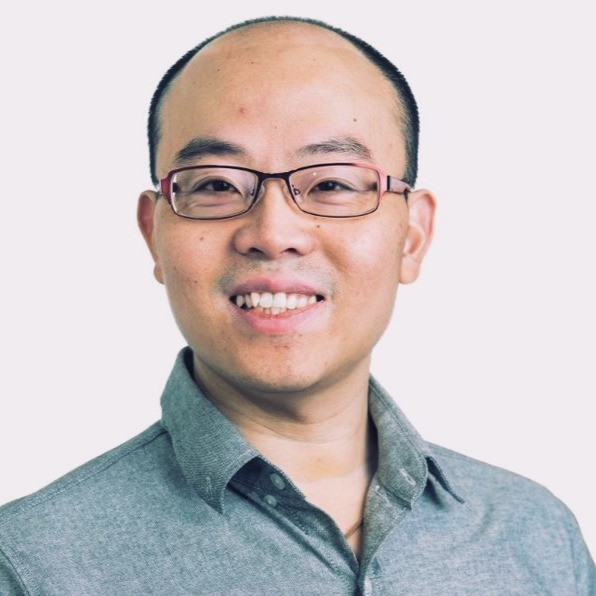}}]{Tanaboon Tongbuasirilai}
received his Ph.D. degree in Media and Information Technology in 2023 and M.Sc. in Advanced Computer Graphics from Link\"{o}ping University, Sweden. He received B.Sc. in Mathematics from Mahidol University, Thailand. His research interests lie in the multiple intersections of computer graphics, image processing, machine learning and sparse representation.
\end{IEEEbiography}

\begin{IEEEbiography}[{\includegraphics[width=1in,height=1.25in,clip,keepaspectratio]{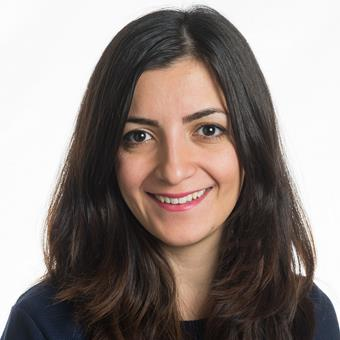}}]{Saghi Hajisharif}
is a Principal Research Engineer at the computer graphics and image processing (CGIP) lab, Link\"{o}ping University, Sweden. Her expertise includes visual machine learning, synthetic data generation, high dynamic range imaging, photo-realistic rendering, compressive imaging, and sparse modeling. She completed her Ph.D. and M.Sc. degrees in computer graphics and image processing at Linköping University in 2020 and 2013, respectively, and her B.Sc. degree in computer science at Amirkabir University of Technology in 2009. Previously, she served as a Postdoctoral Researcher at Linköping University and conducted research as a Visiting Researcher at the Institute for Creative Technologies, University of Southern California.
\end{IEEEbiography}

\begin{IEEEbiography}[{\includegraphics[width=1in,height=1.25in,clip,keepaspectratio]{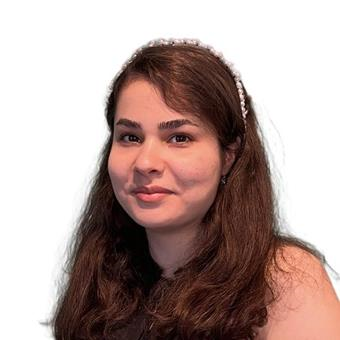}}]{Behnaz Kavoosighafi}
is a Ph.D. candidate in Computer Graphics and Image Processing at Linköping University, Sweden. She received an M.Sc. in Digital Electronics Systems from Shahid Beheshti University, Iran, in 2020 and holds a B.Sc. in Electrical Engineering from K. N. Toosi University of Technology, Iran. Her research interests lie in the domain of computer graphics, specifically in appearance capture and modeling.
\end{IEEEbiography}

\begin{IEEEbiography}[{\includegraphics[width=1in,height=1.25in,clip,keepaspectratio]{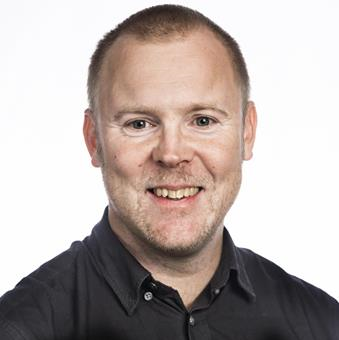}}]{Jonas Unger}
is a professor in computer graphics at Linköping University in Sweden. Unger obtained his PhD in 2009 and became full professor in 2018. Since 2009, he is leading the Computer Graphics and Image Processing group at the department for Science and Technology. Combining sensors, simulation and machine learning, Unger and his group currently pursue research towards new theory and technology for computational imaging by fusing computer graphics, vision and sensors with human perception and machine learning to capture, model, analyze, and synthesize visual aspects of the world in novel ways.  
\end{IEEEbiography}

\vfill

\end{document}